\documentclass[prl,aps,floatfix,10pt,twocolumn,nofootinbib,superscriptaddress]{revtex4-1}

\usepackage[T1]{fontenc}
\usepackage[utf8]{inputenc}
\usepackage{microtype}
\usepackage{array}
\usepackage{graphicx}
\usepackage{dcolumn}
\usepackage{bm}
\usepackage{amsmath}
\usepackage{amsfonts}
\usepackage{amssymb}
\usepackage{amstext}   
\usepackage{amsthm}
\usepackage{array}
\usepackage[english]{babel}
\usepackage{makecell}
\usepackage{dsfont}
\usepackage{nameref}
\usepackage{color}
\usepackage{float}
\usepackage[outdir=./]{epstopdf}
\usepackage{nameref}
\usepackage{subfloat}
\usepackage[colorlinks = true, citecolor = blue, urlcolor =cyan]{hyperref}
\usepackage[figure]{hypcap}
\usepackage{multirow}
\usepackage{makecell}
\usepackage{xcolor}
\usepackage{mathtools}
\usepackage{mathdots}
\usepackage[notransparent]{svg}
\usepackage{todonotes} 
\newcolumntype{C}{>{$}c<{$}}
\AtBeginDocument{
\heavyrulewidth=.08em
\lightrulewidth=.05em
\cmidrulewidth=.03em
\belowrulesep=.65ex
\belowbottomsep=0pt
\aboverulesep=.4ex
\abovetopsep=0pt
\cmidrulesep=\doublerulesep
\cmidrulekern=.5em
\defaultaddspace=.5em
\tabcolsep=7pt
}
\usepackage{booktabs}
\usepackage{verbatim}
\usepackage{ragged2e}

\definecolor{emerald}{rgb}{0.07, 0.53, 0.03}

\newcommand{\ket}[1]{\ensuremath{\vert|#1\rangle}}

\usepackage{braket}

\usepackage{amsbsy}

\newcommand{\tx}[1]{\text{#1}}
\newcommand{\fref}[1]{Fig.~\ref{#1}}

\renewcommand{\eqref}[1]{(Eq.~\ref{#1})}

\usepackage{nicefrac} 
\usepackage[normalem]{ulem} 
\usepackage{soul}   
\setstcolor{red} 
\setul{}{1.5pt}

\usepackage{subfiles}

\begin{document}

\title{The Pound-Drever-Hall Method for Superconducting-Qubit Readout}

\author{Ibukunoluwa Adisa}
\affiliation{Department of Physics, University of Maryland, College Park, MD 20742, USA}
\affiliation{Joint Quantum Institute, NIST/University of Maryland, College Park, Maryland 20742 USA}

\author{Won Chan Lee}
\affiliation{Department of Physics, University of Maryland, College Park, MD 20742, USA}
\affiliation{Joint Quantum Institute, NIST/University of Maryland, College Park, Maryland 20742 USA}

\author{Kevin C. Cox}

\affiliation{Department of Physics, University of Maryland, College Park, MD 20742, USA}
\affiliation{DEVCOM Army Research Laboratory, 2800 Powder Mill Rd, Adelphi MD 20783, USA}

\author{Alicia J.  Koll\'ar}
\affiliation{Department of Physics, University of Maryland, College Park, MD 20742, USA}
\affiliation{Joint Quantum Institute, NIST/University of Maryland, College Park, Maryland 20742 USA}
\affiliation{Maryland Quantum Materials Center, Department of Physics, University of Maryland, College Park, MD 20742, USA}

\preprint{APS/123-QED}

\begin{abstract}
Scaling quantum computers to large sizes requires the implementation of many parallel qubit readouts.
Here we present an ultrastable superconducting-qubit readout method using the multi-tone self-phase-referenced Pound-Drever-Hall (PDH) technique, originally developed for use with optical cavities.
In this work, we benchmark PDH readout of a single transmon qubit, 
using room-temperature heterodyne detection of all tones to reconstruct the PDH signal.
We demonstrate that PDH qubit readout is insensitive to microwave phase drift, displaying $0.73^\circ$ phase stability over 2 hours, and capable of single-shot readout in the presence of phase errors exceeding the phase shift induced by the qubit state.
We show that the PDH sideband tones do not cause unwanted measurement-induced state transitions for a transmon qubit, leading to a potential signal enhancement of at least $14$~dB.

\end{abstract}

\maketitle


\begin{figure}[t]
\centering
    \includegraphics[width=0.48\textwidth]{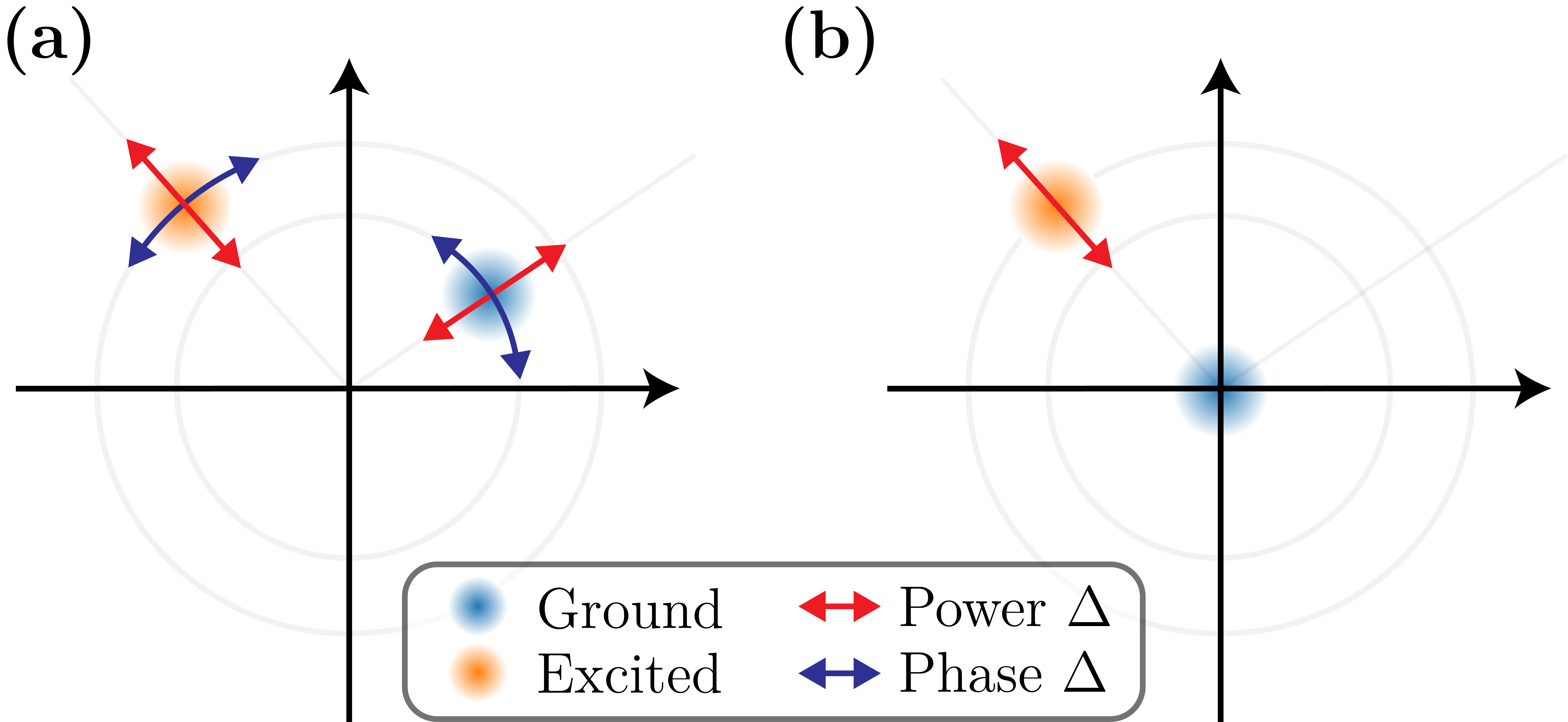}
    \vspace{-0.2cm}
    \caption{
    \textit{Schematic of drift sensitivities for conventional and PDH readout of transmon qubits.} 
    Schematic IQ-plane representation of measurement outcomes corresponding to the ground (blue) and the excited states (orange) of a transmon.
    (a) For heterodyne readout,  the measurement signals are sensitive to both the power (red arrow) and the phase (blue arrow) of the measurement tone. 
    (b) In self phase-referenced PDH readout, the signal is independent of phase drift.}
    \label{fig:readoutSchematics}
\end{figure}

High-fidelity readout of superconducting circuits has enabled transformational demonstrations in quantum science including creation of highly-entangled states~\cite{mooney2019entanglement, huang2026entanglement}, quantum error correction ~\cite{ofek2016QEC, zhao2022QEC, sivak2023QEC, google2025QEC}, and quantum processors with up to 1000 qubits~\cite{IBM1000qubits}.  As quantum systems scale further in performance and size, the requirements on the qubit readout schemes will scale proportionately. Scaling this hardware to tens of thousands of qubits while lowering the cost per channel and improving the performance is a nontrivial challenge. 

Measurement of superconducting qubits \cite{Blais:CavityQED, quantumengineersguide, Blais:CircuitQED, wallraff2005dispersiveReadout, Koch:transmon} is traditionally carried out using dispersive readout, in which the qubit state is inferred from a heterodyne measurement of the state-dependent frequency shifts on a host resonator \cite{Blais:CavityQED, quantumengineersguide, wallraff2005dispersiveReadout, Blais:CircuitQED}. This method has achieved readout speeds below $100$ ns and fidelities above 99.7$\%$ \cite{fastReadoutWallraff, fastReadoutNakamura, readout2017Wallraff}.
However, heterodyne-based dispersive readout is reliant on high-performance clocking of the microwave generators, AWGs, and digitizers, and generally requires rolling background calibration to compensate for slow drift. In contrast, the highest performance optical-frequency measurements often rely on the self-phase-referenced  Pound-Drever-Hall (PDH) method \cite{poundDreverHall1983laser, black2001PDHoverview} for locking lasers and cavities. PDH readout is a critical enabling technique for the world's most precise atomic clocks, routinely achieving optical resonator readout with fractional frequency stability better than one part in $10^{16}$ \cite{robinson2019stabilization,JunYePDH2024, parke2025stabilization}.  
Here we show that the intrinsically phase-stable PDH technique that originated in the optical frequency domain, and has been used for fast characterization of microwave resonators~\cite{PoundLocking_resonatorsCharacterization, PoundLocking_Qmeasurements, ZurichPDH}, can be transported to the quantum microwave regime, where it gives rise to improvements for superconducting qubit readout, in terms of long-time stability, scaling complexity, and signal amplification.  The PDH-based superconducting-qubit readout method we present here \emph{is not} reliant on microwave phase synchronization, and has the potential to improve scalability of superconducting-qubit readout, eliminating the need for rolling reference calibration and potentially reducing the cost per readout channel.

The measured signal of dispersive heterodyne readout after amplification, downconversion, and digital demodulation, takes the form \cite{quantumengineersguide}
\begin{equation} \label{eq:homodynereadout}
    E_H\propto \bar{E}_{0}\bar{E}_{\ell}\cos{[\phi_{0}(\omega_{0};\omega_{\ell})]},
\end{equation}
where $\bar{E}_{0}$ is the carrier amplitude, $\bar{E}_{\ell}$ is the local oscillator (LO) amplitude, and $\phi_{0}(\omega_{0};\omega_{\ell})$ is the measured phase of the readout signal. 
In conventional heterodyne readout, the measured phase includes not only the qubit-dependent phase but also contributions from the microwave sources and dispersion in the signal path such that 
\begin{align} \label{eq:heterodyne_phase}
    \phi_{0}(\omega_{0};\omega_{\ell}) = \phi_{|q\rangle}(\omega_{0}) + \bar{\phi}_{0}(\omega_{0}) - \bar{\phi}_{\ell}(\omega_{\ell})  + \phi_{\tx{path}}(\omega_{0})
\end{align}
where $\phi_{|q\rangle}(\omega_{0})$ is the qubit-dependent phase, $\bar{\phi}_{0}(\omega_{0})$ and $\bar{\phi}_{\ell}(\omega_{\ell})$ are carrier and LO phase contributions, and $\phi_{\tx{path}}(\omega_{0})$ is the phase accrued along the propagation path.
Fluctuations in $\bar{\phi}_0 - \bar{\phi}_\ell$ depend on the quality of the sources and reference locks, and can be eliminated by modulating a single generator to produce both the measurement and LO tones. However,  path-length fluctuations on signals at $\omega_{0}$ (e.g., from thermal drifts) remain a source of phase instability even with high-performance references. 
The resulting phase fluctuations are illustrated in the IQ-plane representation of qubit readout shown in \fref{fig:readoutSchematics}(a).
While such fluctuations can be removed through rolling calibration, a natively insensitive method should be more scalable.

Here we explore an approach for detecting qubit-state-dependent frequency shifts on the resonator based on the PDH method. The readout signal in this multi-tone interrogation scheme consists of three copropagating tones: the carrier ($\omega_{0}$), the upper sideband ($\omega_{+} = \omega_0 + \omega_m$), and lower sideband ($\omega_{-} = \omega_0 -\omega_m$). 
For phase stability, these three tones are produced from a single source, usually via phase modulation 
at $\omega_m$, and the PDH readout signal, derived from the beat between the carrier and sidebands, takes the form 
\begin{align} \label{eq:pdh_error_signal_main}
    \nonumber \epsilon_{\tx{Q}} &= E_{0}E_{+}\sin{\left[\phi_{0}(\omega_{0})- \phi_{+}(\omega_{+})\right]} \\
    & \quad + E_{0}E_{-}\sin{\left[\phi_{0}(\omega_{0}
    ) - \phi_{-}(\omega_{-})\right]},
\end{align}
where $E_{(0,+,-)}$ and $\phi_{(0,+,-)}$ are the amplitudes and phases of the carrier and sidebands.
Note: we use the convention that $\phi_{-}(\omega_{-})$ denotes the \emph{absolute} phase of the lower sideband, which is natural for microwave detection but differs from the standard optical notation by $\pi$ (see Sec. S2B in the Supplemental Material \cite{SupplementalMaterial} for details).
In this approach, the signal depends only on differential phases between the carrier and the sidebands, which traverse the same path. Since the sidebands are derived from the same source as the carrier, generator phase errors are common-mode and largely suppressed. 
Furthermore, unlike in conventional heterodyne readout, where path-length fluctuations comparable to the \emph{carrier} wavelength produce significant phase shifts, in PDH, only fluctuations comparable to the much longer \emph{modulation} wavelength matter. 
Thus, the PDH approach offers an intrinsically phase-stable self-referenced method of detecting the qubit-state-dependent frequency of a microwave resonator, as shown schematically in \fref{fig:readoutSchematics}(b). 

In general, large $E_0$ improves readout SNR and speed; however, for weakly anharmonic qubits, such as transmons, large $E_0$ can also induce unwanted excitations \cite{Blais:transmonionization, YaleMIST_fullCharacterization, YaleMIST_highFrequency}, a process known as measurement-induced state transitions (MIST) \cite{google:MISTbeyondRWA, google:MISTwithinRWA} or transmon ionization \cite{Blais:ionizationdynamics, Blais:transmonionization}. 
The PDH technique can potentially mitigate this limitation through intrinsic heterodyne gain, because the PDH signal \eqref{eq:pdh_error_signal_main} scales as $E_{0}E_{\pm}$. As a result, for sufficiently strong sidebands ($E_{\pm}$), the PDH signal can be much larger for fixed $E_0$, and can overcome technical noise without adding MIST.

While the PDH technique has been highly successful in the optical regime, full implementation of it in the microwave regime has been limited, primarily due to the lack of high-performance square-law detectors to generate and measure the beatnotes. In this work, we implement a PDH-style multi-tone interrogation of superconducting qubits through simultaneous heterodyne detection of the three tones. Our implementation demonstrates two key features of PDH: (i) the intrinsic phase stability of differential readout, and (ii) the absence of sideband-induced MIST for large sideband offsets.   

\begin{figure}[t]
\centering
    \includegraphics[width=0.48\textwidth]{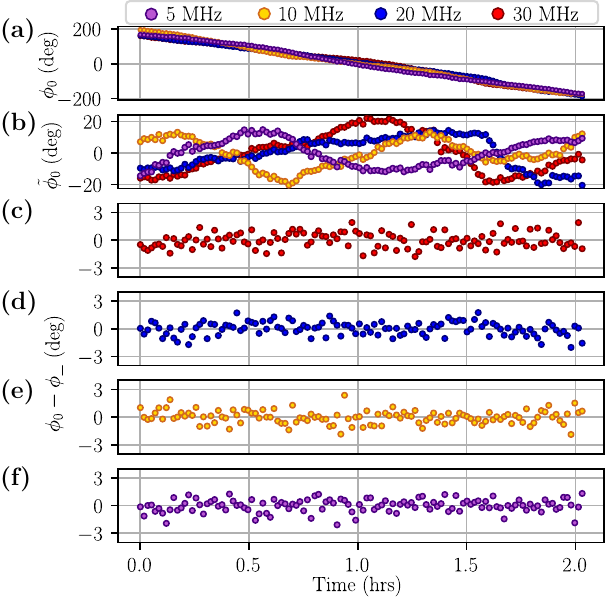}
    \vspace{-0.2cm}
    \caption{\textit{Long-term phase stability of PDH readout.}
    Average carrier and sideband phases over $1000$ consecutive single shots are monitored over a period of $2$ hours for sideband detunings from $5-30$~MHz.
    (a) Raw carrier phase $\phi_{0}$, showing roughly $400^\circ$ drift due to a frequency offset between the carrier and the LO.  
    (b) Non-linear residual fluctuations $\tilde{\phi}_0$ after subtracting the linear drift observed in (a). 
    (c-f) Fluctuations of the PDH differential phase $\phi_{0}$ - $\phi_{-}$ versus sideband detuning. This phase difference displays an RMS phase noise of $0.73^\circ$ due to effective cancellation of common-mode phase noise and drift. Mean phase removed from all data.
    }
    \label{fig:PhaseStabilityReadout}
\end{figure}

\textit{Phase Stability}.--- We investigate the long-term phase stability of the two readout approaches by performing PDH-style interrogation. We independently digitize all three tones, comparing the carrier phase, which determines the conventional heterodyne readout signal, with the differential carrier-sideband phase, which determines the PDH signal.
These measurements, shown in \fref{fig:PhaseStabilityReadout}, illustrate the sensitivity of heterodyne measurement and the robustness of PDH readout.
The carrier frequency is fixed near resonance for all datasets, and the modulation frequency varied from $5$~MHz to $30$~MHz to explore the performance at different sideband offsets. 
The phases are averaged over 1000 back-to-back single shots and sampled stroboscopically over a period of 2 hours.
The carrier phase $\phi_{0}$ (\fref{fig:PhaseStabilityReadout}(a)) exhibits a linear drift of $\Delta \phi_{0} = 400^\circ$ over that period, due to a frequency mismatch of $\Delta f \approx \Delta \phi_{0}/(360^\circ * 7600) \approx 154~\mu$Hz between the RF and LO generators, i.e. a relative precision error of $2.27e^{-14}$ at a carrier frequency of 6.7 GHz.
Although such drift can be suppressed with higher-performance instrumentation and locking schemes (achieving sub $\mu$Hz offsets and sub-degree drifts below the qubit state phase separation), it illustrates the susceptibility of heterodyne readout to clocking errors. 
To highlight contributions that are not simply from frequency offsets, the linear component of each trace is subtracted, revealing non-linear fluctuations $\tilde{\phi}_0$ of up to $\pm 20^\circ$ (\fref{fig:PhaseStabilityReadout}(b)), likely caused by temperature drifts or residual phase-locking errors. While improved lab stability or rolling calibration can mitigate these effects, these results emphasize the fundamental sensitivity of heterodyne readout to \emph{both} clocking errors and laboratory environment.

\begin{figure*}[ht!]
\centering
    \includegraphics[width=0.98\textwidth]{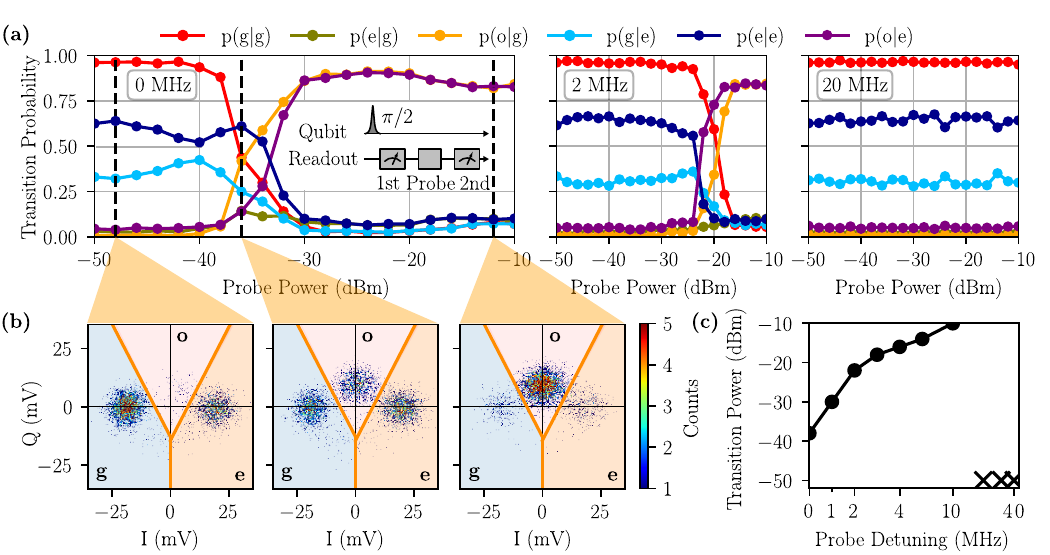}
    \vspace{-0.2cm}
    \caption{\label{fig:QNDness}
    \textit{MIST in PDH readout.}
    MIST due to additional tones during readout is diagnosed using two sequential measurements with a probe tone applied in between (a-inset). 
    Outcomes for each measurement are classified as ground $\ket{g}$, excited $\ket{e}$, or other $\ket{o}$. 
    (a) Conditional probabilities versus probe power and detuning.
    Rapid increases of $p(o|g)$ and $p(o|e)$ indicate MIST above a critical power and breakdown of QND readout.
    (b) IQ histograms for sample $2^{\mathrm{nd}}$ measurements from the $0$~MHz detuning data in (a). Decision boundaries for $\ket{g}$ (blue), $\ket{e}$ (orange), and $\ket{o}$ (red) are shown in gold. 
    (c) Semi-log plot of the transition power at which probe-induced MIST becomes dominant. We empirically define the transition power as the point at which the increase of $p(o|g)$ or $p(o|e)$ exceeds $4\, \%/$dB.
    At sufficiently large detunings, no observable probe-induced MIST transition occurs ($\rm{X}$ markers), even with a sideband at $28$~dBc, corresponding to more than $14$~dB of heterodyne gain during PDH error-signal detection.
    }
\end{figure*}

In contrast, all these sensitivities are removed with PDH readout, in which the sideband is used as a phase reference. The measured differential phases between the carrier and the lower sideband, for sideband detunings between $5$~MHz and $30$~MHz, are shown in \fref{fig:PhaseStabilityReadout}(c-f) (see Sec. S2 in the Supplemental Material~\cite{SupplementalMaterial} for details of the phase convention and extraction procedure). Despite the ~$400^\circ$ absolute drift in $\phi_{0}$, the differential phase remains stable with RMS fluctuations of only ~$0.73^\circ$, averaged over modulation frequencies.
The measured phase stability is independent of sideband detuning even though cryogenic coax typically displays non-negligible dispersion over comparable frequency ranges. This enables the use of large detunings (> $20$ MHz), which are expected to be optimal for qubit readout since they allow for large heterodyne gain without added MIST (see \fref{fig:QNDness}) and short detection windows.

\begin{figure}[t!]
\centering

    \includegraphics[width=0.48\textwidth]{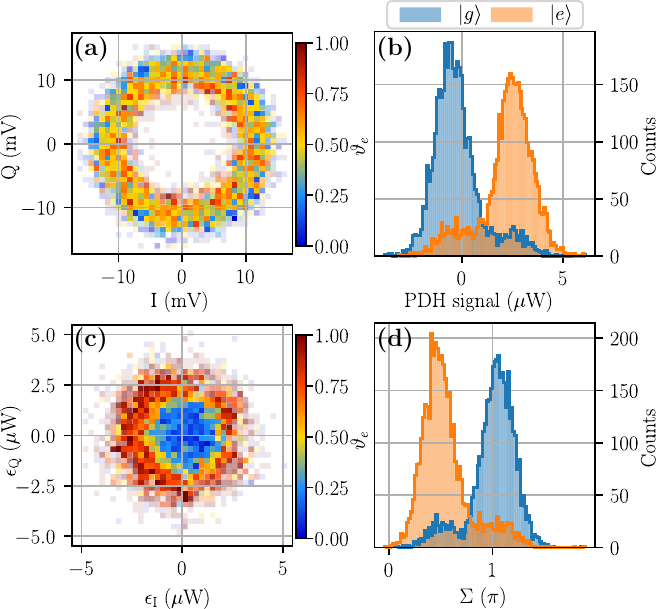}
    \vspace{-0.2cm}
    \caption{
    \textit{Single-shot phase stability of PDH readout.} 
    Qubit readout with free running generators for the qubit prepared either in $\ket{g}$ or $\ket{e}$. (a) Heterodyne IQ distribution for the carrier. Loss of phase coherence collapses the two states into overlapping IQ distributions.
    The color map indicates $\vartheta_e$, the fraction of events corresponding to preparation in $\ket{e}$. 
    Transparency is reduced for pixels with less than 10 counts, where $\vartheta_e$ is noisy.
    (b)  PDH readout with the same data, using the optimal combination of real and imaginary parts ($\epsilon_I$ and $\epsilon_Q$), preserves clear state separation, demonstrating robustness against carrier phase errors. 
    Imperfect state preparation results in a small fraction of mislabeled events and overlapping pedestals in the histograms for $\ket{g}$ and $\ket{e}$
    (see Sec. S3 in the Supplemental Material~\cite{SupplementalMaterial} for details).
    (c) PDH readout with both carrier phase errors and RF timing errors. The latter rotate the real and imaginary parts of the PDH signal, compromising readout fidelity. 
    (d) Scissors-phase readout, defined as $\Sigma = 2\phi_{0} - (\phi_{-}+\phi_{+})$, remains robust against the stated phase and timing errors maintaining state discrimination.
    }
    \label{fig:heterodyne_pdh_sigma}
\end{figure}

In addition to the long-term phase stability measurements above, we demonstrate the single-shot phase stability of PDH and show that a usable signal can be recovered even in the presence of carrier-phase errors comparable to $2 \pi$.
In this work, we use a reconstructed PDH signal obtained from simultaneous heterodyne detection of the carrier and the sidebands using a parametric amplifier \cite{yurke1989JPA, siddiqi2004JPA, mutus2013JPA} (see Sec. S8C in the Supplemental Material~\cite{SupplementalMaterial} for a comparison to conventional heterodyne and Sec. S2E for discussion of the reconstruction process and a comparison to direct generation of the PDH signal).
Measurements are performed with the qubit prepared in both $\ket{g}$ and $\ket{e}$ states and the carrier tuned such that the heterodyne IQ readout separation lies primarily in phase \cite{quantumengineersguide, practicalGuide_gao2021}.
Carrier-phase errors are deliberately induced by unlocking the microwave generators, and the heterodyne measurements corresponding to the two qubit states, shown in \fref{fig:heterodyne_pdh_sigma}(a) are essentially indistinguishable. 
In contrast, PDH readout, shown in \fref{fig:heterodyne_pdh_sigma}(b), preserves separability between the qubit states due to \emph{carrier-sideband} phase coherence. 

A further advantage of PDH readout via simultaneous heterodyne detection of the carrier and sidebands is the possibility of reconstructing more general phase combinations not directly obtainable with a photodiode. 
In particular, the phase combination $\Sigma = 2\phi_{0} - (\phi_{-} + \phi_{+})$, which we term the scissors phase in analogy to the normal modes of ion traps \cite{exploringTheQuantum}, 
is orthogonal to the correlated phase shifts induced by errors in both the modulation and detection timing, as well as the generator  phase (see Sec. S4 in the Supplemental Material~\cite{SupplementalMaterial} for details).
While such timing errors are uncommon in optical PDH setups, where a single radio-frequency source drives both modulation and demodulation, microwave setups often have independent sources.
To demonstrate that $\Sigma$ is robust against modulation and detection errors which compromise PDH, we examine single-shot readout while deliberately inducing timing offsets on \emph{all} components including the carrier generator, the modulator, and the digitizer. 
Under these conditions, PDH yields low-fidelity readout (\fref{fig:heterodyne_pdh_sigma}(c)) due to quadrature mixing, but the scissors phase, shown in  \fref{fig:heterodyne_pdh_sigma}(d), maintains separation of ground and excited states.

\textit{Intrinsic Heterodyne Gain}.--- Because the PDH signal comes from the beats between the carrier and the sidebands, PDH qubit readout can experience effective amplification analogous to mixing with a strong LO if the sidebands are strong compared to the carrier. This intrinsic heterodyne gain can mitigate added noise in down-stream amplifiers and could allow PDH readout with a cryogenic $|E|^2$ detector to achieve higher readout SNR than conventional heterodyne. Here we demonstrate that sidebands strong enough to achieve more than $14$~dB of heterodyne gain can be applied without inducing additional MIST.

To quantitatively benchmark how much sideband amplitude $E_\pm$ the qubit can tolerate, we evaluate the state transitions 
versus measurement-tone frequency and power
in a two-measurement scheme modified from Ref.~\cite{AngelaKouMIST}, shown schematically in the inset of \fref{fig:QNDness}(a). 
An initial $\pi/2$ pulse is applied to the transmon qubit to prepare it in an equal superposition of the ground and excited states.  
For each configuration, $5000$ single shots are performed. Each shot consists of two consecutive $2$-$\mu$s-long heterodyne measurements separated by $8\ \mu$s, and 
between these, a probe is applied that replicates the effect of a measurement tone with variable-frequency and variable-power. 
Repeated application of an ideal QND measurement will always yield the same outcome, i.e. the conditional probability $p(j|j) = 1$. Due to imperfect measurement fidelity, some non-zero transition probabilities ($p(j|i) \neq 0$ for $i \neq j$) are always present, but the effect of the probe tone alone can be determined by examining the power and frequency dependence of the transition probabilities, and shown in \fref{fig:QNDness}(a).

To determine the individual measurement outcomes, the IQ plane is divided into three regions, corresponding to the two qubit states, as well as a designated ``other'' outcome $\ket{o}$, corresponding to all higher-energy levels of the transmon. Decision boundaries are determined using a reference resonant probe power for which the measurement outcomes are trimodal due to the onset of strong MIST (see Sec. S6 in the Supplemental Material~\cite{SupplementalMaterial} for details on the decision boundary determination). Sample IQ histogram plots from the second measurement at $0$~MHz detuning are shown in \fref{fig:QNDness}(b), along with the decision boundaries and color-coded outcome regions. All IQ points falling within the same region are assigned to the same state.

The power and detuning dependence of the conditional measurement probabilities $p(j|i)$ are shown in \fref{fig:QNDness}(a). 
When the probe detuning is zero, the possible range of applied powers is severely limited by the onset of added MIST. 
Significant increases in the increase of $p(o|g)$ or $p(o|e)$ above $4\, \%/$dB are observed at approximately $-38$~dBm. 
See left-hand panel in \fref{fig:QNDness}(a). 
Above this point, heterodyne readout would be compromised, and the increase of the measurement-induced $\ket{o}$ population with increasing probe power can be clearly seen in the IQ histogram shown in the third plot of \fref{fig:QNDness}(b).

In the case of PDH readout, however, additional signal amplitude can be gained by leveraging strong off-resonant sidebands.
When the detuning of the probe from the carrier frequency exceeds $20$~MHz (approximately $33$ times the resonator linewidth), no observable signature of added MIST is found within the range of applied probe powers (see Sec. S6 in the Supplemental Material~\cite{SupplementalMaterial} for additional details and detuning dependence).
Thus, the PDH readout scheme can tolerate at least $28$~dBc more power in the sideband at large detunings, corresponding to $14$~dB of potential heterodyne gain for the PDH error signal, without added MIST (see Sec. S8 subsection B3 in the Supplemental Material~\cite{SupplementalMaterial}).

In conclusion, we have demonstrated that the PDH technique can be successfully extended to the quantum microwave domain as a readout scheme for superconducting qubits, preserving robust phase stability even in the presence of strong dispersion and exhibiting the potential for significant intrinsic heterodyne gain. 
Our approach, implemented without a cryogenic $|E|^2$ detector through PDH-style multi-tone interrogation, yields substantial gains in intrinsic phase stability, dramatically reducing both the technical requirements and potential cost per channel of readout electronics, and making the PDH method particularly promising for large-scale integration with superconducting qubits.
Ultimately, the development of a Josephson-junction-based cryogenic $|E|^2$ detector will enable implementation of PDH with intrinsic heterodyne gain, which could surpass conventional heterodyne measurements by achieving higher SNR for fixed resonant probe power.
This prospect motivates the development of a new generation of Josephson devices specifically optimized to detect and amplify PDH beatnotes, rather than for linear amplification of the carrier field, paving the way to ultrastable, scalable, and cost-efficient superconducting-qubit readout architectures.

\begin{acknowledgments}

This project was supported by ARL Grants Nos. W911NF-24-2-0107, W911NF-19-2-0181, and W911NF-17-S-0003. IA, WCL, and AK received additional support from NSF Grant No. PHY2047732, the Sloan Fellowship Program, AFOSR Grant No. Grant No. FA9550-21-1-0129, the NSF QLCI grant OMA-2120757, and the University of Maryland. Research at UMD was also supported by the Maryland Quantum Materials Center.
 
We thank Jun Ye, Martin Ritter, and Nathan Schine for helpful comments and discussions, and we thank Jim Phillips and the McRae Group for helpful conversations regarding fast characterization and Q-factor measurements of superconducting resonators using PDH.

\end{acknowledgments}

\bibliographystyle{apsrev4-2}
\bibliography{refs.bib}

\end{document}


\title{Supplemental Material For: The Pound-Drever-Hall Method for Superconducting-Qubit Readout}

\author{Ibukunoluwa Adisa}
\affiliation{Department of Physics, University of Maryland, College Park, MD 20742, USA}
\affiliation{Joint Quantum Institute, NIST/University of Maryland, College Park, Maryland 20742 USA}

\author{Won Chan Lee}
\affiliation{Department of Physics, University of Maryland, College Park, MD 20742, USA}
\affiliation{Joint Quantum Institute, NIST/University of Maryland, College Park, Maryland 20742 USA}

\author{Kevin C. Cox}
\affiliation{Department of Physics, University of Maryland, College Park, MD 20742, USA}
\affiliation{DEVCOM Army Research Laboratory, 2800 Powder Mill Rd, Adelphi MD 20783, USA}

\author{Alicia J.  Koll\'ar}
\affiliation{Department of Physics, University of Maryland, College Park, MD 20742, USA}
\affiliation{Joint Quantum Institute, NIST/University of Maryland, College Park, Maryland 20742 USA}
\affiliation{Maryland Quantum Materials Center, Department of Physics, University of Maryland, College Park, MD 20742, USA}

\preprint{APS/123-QED}

\maketitle

\setcounter{figure}{0}
\setcounter{equation}{0}
\renewcommand{\figurename}{Figure}
\renewcommand{\thefigure}{S\arabic{figure}}
\renewcommand{\thetable}{S\arabic{table}}
\renewcommand{\theequation}{S\arabic{equation}}

\renewcommand{\thesection}{S\arabic{section}}

\section{Notation and Conventions}
Understanding the Pound-Drever-Hall (PDH) and heterodyne detection techniques requires keeping track of the amplitude and phase relationships between multiple AC signals at different frequencies. This Appendix establishes the unified notation used throughout this work. For consistency of notation, we adopt an optics-like notation, using symbols based on electric-field representations even when discussing microwave signals. In the microwave case, these quantities correspond physically to voltages in coaxial cables, but the single notation defined here is used for both optical and microwave cases. We summarize the notation below. 

\begin{itemize}
    \item Barred quantities (e.g., $\bar{E}$ and $\bar{\phi}$) denote the absolute quantities produced by the source generator. For instance, $\bar{\phi}$ represents the absolute phase produced by the source.

    \item Unbarred quantities (e.g., $E$ and $\phi$) denote measured or detected quantities, typically defined relative to the local oscillator in the detection chain. 

    \item Complex fields are written in boldface ($\boldsymbol{E}$), while time-dependent fields are expressed in calligraphic symbol $\mathcal{E}(t)$. The real-valued amplitudes of these fields are written in standard fonts ($E$). Thus, a complex time-dependent field $\boldsymbol{\mathcal{E}}(t) = \boldsymbol{E} e^{i \omega t} = E e^{i \phi} e^{i \omega t}$.

\end{itemize}

These conventions will be used consistently throughout the derivations and experimental discussions that follow. 

\section{Qubit Readout}\label{app:qubitreadout}
In this Appendix, we will briefly review the dispersive coupling mechanism used to link the state of a superconducting qubit to the frequency of a microwave resonator~\cite{Blais:CircuitQED, wallraff2005dispersiveReadout} and describe how this qubit-state-dependent frequency shift can be extracted using either heterodyne or PDH detection. We will review the signal processing involved in each detection method and their sensitivities to experimental imperfections.

\subsection{Dispersive Readout of Superconducting Qubits}\label{app:dispersivereadout}

In the circuit quantum electrodynamics (cQED) framework \cite{Blais:CircuitQED, wallraff2005dispersiveReadout, Koch:transmon}, quantum information processing is achieved by coupling superconducting qubits to microwave resonators that facilitate readout. In a typical simple system, a qubit is coupled to a single mode of the readout resonator with an interaction that is described by the Jaynes-Cummings Hamiltonian \cite{steck2007quantum} 
\begin{equation} \label{eq:Jaynes-Cummings}
    H_{\tx{JC}} = \omega_{r}\left(a^{\dagger}a\right) - \frac{\omega_{q}}{2}\sigma_{z} + g(a\sigma^{+} + a^{\dagger}\sigma^{-}),
\end{equation}
where $\omega_{r}$ and $\omega_{q}$ denote the resonator frequency and qubit transition frequency, respectively, and $g$ is the qubit-resonator coupling rate. The operators $a\left(a^{\dagger}\right)$ are the annihilation (creation) operators of the resonator, and $\sigma^{-}\left( \sigma^{+}\right)$ are the lowering and raising operators for the qubit. For simplicity, we describe the transmon qubit as a two-level system, a full treatment, including the effects of higher energy levels, can be found in \cite{Koch:transmon}. When the qubit-cavity interaction is in the dispersive regime, i.e. when the qubit-cavity detuning $\Delta_{q-r} = \omega_{q} - \omega_{r}$ is much larger than the coupling rate $\left( \Delta_{q-r} \gg g\right)$ the Jaynes-Cummings Hamiltonian in \eqref{eq:Jaynes-Cummings} can be approximated as \cite{Blais:CircuitQED} 
\begin{equation} \label{eq: dispersive_hamiltonian}
    H_{\tx{disp}} = \left(\omega_{r} - \chi\sigma_{z} \right)a^{\dagger}a - \left(\omega_{q} + \chi \right)\frac{\sigma_{z}}{2},
\end{equation}
where $\chi = g^{2}/\Delta_{q-r}$ is the dispersive shift. In this regime, the qubit does not exchange energy with the readout resonator but instead shifts the resonator frequency by $\pm \chi$ depending on the state of qubit \cite{wallraff2005dispersiveReadout}. This state-dependent frequency shift forms the basis of \emph{dispersive readout}, in which probing the cavity response with a microwave tone near resonance yields a transmitted or reflected signal whose phase and amplitude encode the qubit state. In the ideal limit, this measurement is quantum non-demolition (QND) \cite{Blais:CavityQED}. Comprehensive treatments of the dispersive Hamiltonian, measurement theory, and experimental implementations can be found in \cite{Blais:CircuitQED, quantumengineersguide, wallraff2005dispersiveReadout, Blais:CavityQED}. In practise, dispersive readout is typically implemented with homodyne or heterodyne detection in which the carrier signal is mixed with a stable local oscillator (LO). The latter approach, detailed below, forms the basis of the conventional heterodyne readout.

\subsection{Heterodyne Readout} \label{app:heterodyne_readout}

\begin{figure} [t!]
\centering
    \includegraphics[width=0.8\textwidth]{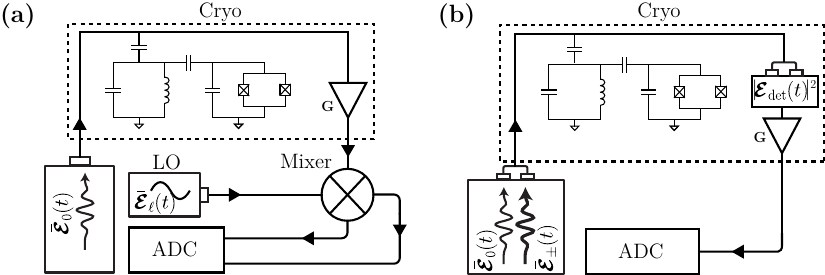}
    \vspace{-0.2cm}
    \caption{
    \textit{Schematic of measurement configurations for conventional heterodyne and PDH readout of transmon qubits.} (a) Conventional heterodyne readout of transmon qubits. A single readout signal, the carrier $\bar{\boldsymbol{\mathcal{E}}}_{0}(t)$, is used to interact with the device in the dilution refrigerator. The output signal from the device is amplified by an amplifier chain with total gain G. The amplified signal is then mixed with a local oscillator signal $\bar{\boldsymbol{\mathcal{E}}}_{\ell}(t)$ and digitally demodulated to produce the final readout signal.
    (b) Ideal Pound-Drever-Hall (PDH) readout of transmon qubit. The readout signal, consisting of the carrier $\bar{\boldsymbol{\mathcal{E}}}_{0}(t)$ and two sidebands $\bar{\boldsymbol{\mathcal{E}}}_{\pm}(t)$, interacts with the device and the output is incident on a power detector which measures the total power $|\boldsymbol{\mathcal{E}}_{\tx{det}}(t)|^{2}$. The resulting beatnote signal, carrying the cavity-induced phase shift, is amplified by the amplifier chain to yield the PDH readout output after digital demodulation.
    }
    \label{fig:readoutSchematics_supple}
\end{figure}
Having outlined the general principles of dispersive readout, we now turn to its conventional realization through heterodyne detection. This method provides simultaneous access to the amplitude and phase of the readout microwave signal and has enabled high-fidelity qubit-state discrimination \cite{quantumengineersguide, Blais:CavityQED, wallraff2005dispersiveReadout}. A typical readout schematic for the conventional heterodyne readout method is shown in \fref{fig:readoutSchematics_supple}(a). A single microwave tone  $\bar{\boldsymbol{\mathcal{E}}}_{0}(t) = \bar{E}_{0}e^{i\left(\bar{\phi}_{0}+\omega_{0} t\right)}$, generated by the microwave source with amplitude $\bar{E}_{0}$, phase $\bar{\phi}_{0}$, and frequency $\omega_{0}$ (chosen near or on resonance with the readout resonator), is sent into the dilution refrigerator. The signal returning from the fridge is modified by the scattering coefficient $\boldsymbol{S}(\omega_{0})$ describing the response of the resonator and propagation along the input-output coax lines. Thus, the outgoing signal is 
\begin{align}
    \nonumber
    \boldsymbol{\mathcal{E}}_{\tx{out}}(t) &= \bar{E}_{0}e^{i\left(\bar{\phi}_{0}+\omega_{0} t\right)} \cdot \boldsymbol{S}(\omega_{0})\\ 
    & = \bar{E}_{0}A(\omega_{0})e^{i\omega_{0}t}e^{i\left( \bar{\phi}_{0} + \phi_{\ket{q}}  + \phi_{\tx{path}}\right)}
\end{align}
where $A(\omega_{0}) = |\boldsymbol{S}(\omega_{0})|$ is the amplitude response at frequency $\omega_{0}$, and $\tx{arg}\left[\boldsymbol{S}(\omega_{0}) \right] = \phi_{\ket{q}} + \phi_{\tx{path}}$ is the total phase shift, which includes the qubit-state-dependent phase shift $\phi_{\ket{q}}$ and the qubit-independent propagation phase $\phi_{\tx{path}}$. After the outgoing signal passes through the amplifier chain with total gain G, at detection, the amplified signal is downconverted with a reference oscillator $\bar{\boldsymbol{\mathcal{E}}}_{\ell}(t) = \bar{E}_{\ell}e^{i\left(\bar{\phi}_{\ell}+\omega_{\ell} t\right)}$ using an analog mixer to transform the high-frequency microwave signal into a low-frequency RF signal that can be sampled correctly by an analog-to-digital converter (ADC). Mixing the signals results in terms at both the sum and difference of the two frequencies $\left( \omega_{0} \pm\omega_{\ell}\right)$, however, the low-frequency terms at the intermediate frequency (IF) $\omega_{\tx{IF}} = \omega_{0} - \omega_{\ell}$ can be isolated using low-pass filters to produce the resulting intermediate component
\begin{align}
    \nonumber
    \boldsymbol{\mathcal{E}}_{\tx{IF}}(t) &= \frac{1}{8}\tx{G}\bar{E}_{0}A(\omega_{0})\bar{E}_{\ell}e^{i\omega_{\tx{IF}}t}e^{i\left( \bar{\phi}_{0} - \bar{\phi}_{\ell} + \phi_{\ket{q}} + \phi_{\tx{path}}\right)} \\
    & = \boldsymbol{I}_{\tx{IF}}(t) + i\boldsymbol{Q}_{\tx{IF}}(t).
\end{align}
Here,
\begin{align}
    \boldsymbol{I}_{\tx{IF}}(t) & = \frac{1}{8}\tx{G}\bar{E}_{0}A(\omega_{0})\bar{E}_{\ell} \cos{\left[\omega_{\tx{IF}}t + \bar{\phi}_{0} - \bar{\phi}_{\ell} + \phi_{\ket{q}} + \phi_{\tx{path}}\right]}, \\
    \boldsymbol{Q}_{\tx{IF}}(t) & = \frac{1}{8}\tx{G}\bar{E}_{0}A(\omega_{0})\bar{E}_{\ell} \sin{\left[\omega_{\tx{IF}}t + \bar{\phi}_{0} - \bar{\phi}_{\ell} + \phi_{\ket{q}} + \phi_{\tx{path}}\right]}
\end{align}
are the downconverted IQ components oscillating at $\omega_{\tx{IF}}$, a frequency small enough to be sampled at high resolution by the ADC but large enough to avoid the 1/f noise at DC. Once digitized, the sampled IQ data are first digitally demodulated by multiplying the signal point-by-point by $\cos{\left( \omega_{\tx{IF}}t\right)}$ and $\sin{\left( \omega_{\tx{IF}}t\right)}$, which shifts the IF signal to DC. The resulting components are then low-pass filtered below $\omega_{\tx{IF}}$ to remove residual terms. The final IQ signal is thus
\begin{align}
    I & = \frac{1}{16}\tx{G}\bar{E}_{0}A(\omega_{0})\bar{E}_{\ell} \cos{\left[ \bar{\phi}_{0} - \bar{\phi}_{\ell} + \phi_{\ket{q}} + \phi_{\tx{path}}\right]}, \\
    Q & = \frac{1}{16}\tx{G}\bar{E}_{0}A(\omega_{0})\bar{E}_{\ell} \sin{\left[ \bar{\phi}_{0} - \bar{\phi}_{\ell} + \phi_{\ket{q}} + \phi_{\tx{path}}\right]}.
\end{align}
The total readout phase  
\begin{align}
    \nonumber
    \phi_{0} &= \arctan{\left(Q/I\right)} \\
    & = \bar{\phi}_{0} - \bar{\phi}_{\ell} + \phi_{\ket{q}} + \phi_{\tx{path}}
\end{align}
of the measured signal consists not only of the qubit-state-dependent phase which is the primary phase of interest, but also other phase contributions from other sources, such as the microwave generators ($\bar{\phi}_{0}$ and $\bar{\phi}_{\ell}$), and the propagation path length ($\phi_{\tx{path}}$).  
Apparent changes in the measured phase can be induced by multiple effects which are unrelated to the qubit. Below, we discuss the dominant effects in several different scenarios: 

\vspace{0.1in}
\noindent \textit{Well-locked generators}: In the ideal case where both carrier and LO generators are high-performance and tightly locked, the carrier–LO phase fluctuations are strongly suppressed, effectively eliminating fluctuations from the generator phases ($\delta\bar{\phi}_{0} - \delta\bar{\phi}_{\ell}$). 
The ideal limit of this is when the carrier and the LO are generated from the same source. This can be done with modulation and filters or using an FPGA-based multi-channel source.
The dominant residual phase-noise channel is then the path-dependent contribution ($\delta \phi_{\tx{path}}$), arising from environmental factors such as temperature fluctuations. Importantly, in conventional heterodyne readout, path-length fluctuations which are comparable to the carrier wavelength produce significant phase shifts, and these residual fluctuations manifest as non-linear phase drift in the readout phase, as seen in Fig. 2(b) (in main text). Therefore, even under optimal locking conditions, conventional heterodyne readout remains intrinsically sensitive to environmental phase drifts.

\vspace{0.1in}
\noindent \textit{Generators locked to the same reference but with relative frequency mismatch}: In practice, different generator models use different phase-locked loop (PLL) configurations to lock to a reference source, which can lead to unwanted frequency offsets even when both the main and LO generators are tightly locked to the reference clock. An example of such a case is shown in Fig.~2(a) in the main text, where the measured phase exhibits a linear drift in time primarily due to the unwanted frequency offset between the carrier and LO generators.

\vspace{0.1in}
\noindent \textit{Free-running or uncorrelated generators}: When the carrier and LO generators are not locked to a common external frequency reference, their phase contributions are uncorrelated. Thus, the measurement of $\bar{\phi}_{0} - \bar{\phi}_{\ell}$ wanders arbitrarily over time, masking all information about the qubit state in the measured heterodyne phase $\phi_{0}$, as shown in Fig.~3(a-i) in the main text. For a qubit-state measurement where the qubit-state information is mainly in the phase quadrature, the resulting IQ-plane distributions collapse into overlapping rings, making state discrimination unreliable, as shown in Fig.~3(a-i) in the main text. 
    
\vspace{0.1in}
In summary, while conventional heterodyne detection has been highly successful in the readout of superconducting qubits, it is intrinsically sensitive to absolute phase drift, and its stability depends critically on both the quality of the external clocking scheme and the physical stability of the microwave signal path. However, using PDH readout, which will be discussed in the next Appendix, robust readout is observed even when the generators are free-running.

\subsection{PDH Readout}
The Pound-Drever-Hall (PDH) technique was first introduced in Ref.~\cite{poundMicrowavestabilization} for stabilizing microwave oscillators using a high-Q cavity as a frequency discriminator, and later extended to optics \cite{poundDreverHall1983laser}, where it has become a key scheme in precision frequency stabilization. In its modern form \cite{black2001PDHoverview}, PDH employs radio-frequency (RF) phase modulation to generate sidebands on a carrier, after which a square-law power detector (e.g., photodiode) is used to create and measure the beatnotes between the carrier and the sidebands. The resulting error signal, which depends only on the differential phase between the carrier and the sidebands, provides a sensitive frequency discriminator largely immune to phase noise. Recently, PDH has been used in the microwave domain for fast high-resolution characterization of superconducting resonators \cite{PoundLocking_resonatorsCharacterization, PoundLocking_Qmeasurements}, enabling accurate real-time tracking of the resonance frequency and quality factor.

In this work, we extend the implementation of PDH in the microwave regime to measurement of superconducting qubits. In this context, the PDH error signal becomes qubit-state-dependent, since the resonator frequency is determined by the qubit state, and provides a highly stable method of inferring the qubit state. Ideal microwave PDH readout approach follows the schematic in \fref{fig:readoutSchematics_supple}(b). The input signal that goes into the dilution refrigerator is a phase-modulated signal consisting of three different tones -- the lower sideband, the carrier, and the upper sideband. Optical PDH is most naturally formulated in terms of the \emph{extra} phases accrued by an ideal phase modulated signal (with equal and opposite sidebands) due to interaction with the cavity \cite{black2001PDHoverview}. However, for our microwave implementation, it is more natural to formulate the problem in terms of the \emph{absolute} phases of the carrier and each sideband. Below, we reproduce the standard treatment of PDH within this convention. 

We define $(\bar{E}_{-}, \omega_{-}, \bar{\phi}_{-})$, $(\bar{E}_{0}, \omega_{0}, \bar{\phi}_{0})$, $(\bar{E}_{+}, \omega_{+}, \bar{\phi}_{+})$ as the (amplitude, frequency, phase) of the lower sideband, carrier, and upper sideband tones generated from the signal generator, respectively. In this notation, the general input signal for PDH consists of 3 independent tones and takes the form
\begin{equation} \label{eq:phase modulation}
    \boldsymbol{\bar{\mathcal{E}}} (t) = \bar{E}_{-}e^{i(\omega_{-}t + \bar{\phi}_{-})} + \bar{E}_{0}e^{i(\omega_{0}t + \bar{\phi}_{0})}+\bar{E}_{+}e^{(i\omega_{+}t + \bar{\phi}_{+})},
\end{equation}
where $\bar{E}_{-} = \bar{E}_{+}$, $\bar{\phi}_{+}=\bar{\phi}_{0}$ and $\bar{\phi}_{-} = \bar{\phi}_{0}+\pi$ for the case of ideal phase modulation. We use the fully general version of input field in \eqref{eq:phase modulation} because it is the most natural for microwave implementation, in which all the three tones are generated and measured individually (see Appendix \ref{sec:TripleDDC}), and because it allows for analysis of phase and amplitude imperfections. Additionally, since the microwave sideband tones are produced via IQ modulation, rather than phase modulation, the amplitudes of the two sidebands can be controlled independently, and without automatic production of higher-order sidebands. A similar approach of directly generating only the three relevant tones for the PDH protocol has also been recently implemented in optical domain \cite{JunYePDH2024} for better control of residual amplitude modulation, unlike in the traditional implementation of PDH \cite{poundDreverHall1983laser, black2001PDHoverview}, where an electro-optic modulator is used to modulate a carrier signal and produces multiple orders of sidebands around the carrier signal.

After the input signal in \eqref{eq:phase modulation} interacts with the resonator, each tone picks up a factor of the scattering coefficient $\textbf{S}(\omega)$ at the appropriate frequency. Thus, the field arriving at the power detector is 
\begin{align} \label{eq:detector_signal}
    \boldsymbol{\mathcal{E}}_{\tx{det}}(t) = \bar{E}_{-}\textbf{S}(\omega_{-})e^{i(\omega_{-}t + \bar{\phi}_{-})} + \bar{E}_{0}\textbf{S}(\omega_{0})e^{i(\omega_{0}t + \bar{\phi}_{0})} + \bar{E}_{+}\textbf{S}(\omega_{+})e^{i(\omega_{+}t + \bar{\phi}_{+})}.
\end{align}
For convenience, we re-write $\bar{E}e^{i\bar{\phi}}\cdot\textbf{S}(\omega) = E(\omega)\cdot e^{i\phi(\omega)}$ for each component, where $E(\omega)$ and $\phi(\omega)$ are the total amplitude and total phase of the component \emph{at the detector}.
Thus,
\begin{align} \label{eq:detector_signal2}
    \boldsymbol{\mathcal{E}}_{\tx{det}}(t) = E_{-}(\omega_{-})e^{i\omega_{-}t}e^{i\phi_{-}(\omega_{-})}
    + E_{0}(\omega_{0})e^{i\omega_{0}t}e^{i\phi_{0}(\omega_{0})} 
    + E_{+}(\omega_{+})e^{i\omega_{+}t}e^{i\phi_{+}(\omega_{+})}.
\end{align}
From this point onward, to simplify notation, we drop the explicit frequency dependence of the amplitudes $E_{0,\pm}$ and phases $\phi_{0,\pm}$ on the corresponding frequencies $\omega_{0,\pm}$. The subscripts of $E_{0,\pm}$ and $\phi_{0,\pm}$ implicitly denote their respective frequency components unless otherwise specified. Therefore \eqref{eq:detector_signal2} is rewritten as
\begin{align} \label{eq:detector_signal2_1}
    \boldsymbol{\mathcal{E}}_{\tx{det}}(t) = E_{-}e^{i\omega_{-}t}e^{i\phi_{-}}
    + E_{0}e^{i\omega_{0}t}e^{i\phi_{0}} 
    + E_{+}e^{i\omega_{+}t}e^{i\phi_{+}}.
\end{align}
The output of the detector is the power of the incident signal, given by
\begin{align} \label{eq:power_signal}
    \nonumber
    \mathcal{P}_{\tx{det}}(t) &= \boldsymbol{\mathcal{E}}_{\tx{det}}(t) \cdot \boldsymbol{\mathcal{E}}^{*}_{\tx{det}}(t) \\
    \nonumber
     &= E_{-}^{2} + E_{0}^{2} + E_{+}^{2}~+ \\
     \nonumber
     & \quad 2E_{-}E_{0}\tx{Re}[e^{i(\phi_{-} - \phi_{0})}e^{-i\omega_{m}t}]  + 2E_{0}E_{+}\tx{Re}[e^{i(\phi_{0} - \phi_{+})}e^{-i\omega_{m}t}]~+ \\
     & \quad 2E_{-}E_{+}e^{i(\phi_{+} - \phi_{-})}e^{i2\omega_{m}t},
\end{align}
which contains multiple frequency components: the DC terms due to each individual tone, the terms oscillating at $\omega_{m}$, and a term oscillating at $2\omega_{m}$ from the interaction between the sidebands. Out of these components, the terms oscillating at $\omega_{m}$ are of the most importance for readout purposes because they originate from the interaction of the carrier with each of the sidebands and thus contain the differential phase $\phi_{0}-\phi_{\pm}$ that carries the qubit-state information.  
Expanding \eqref{eq:power_signal},
\begin{align} \label{eq:power_signal2}
    \nonumber
    \mathcal{P}_{\tx{det}}(t) &= E_{-}^{2} + E_{0}^{2} + E_{+}^{2} - 2E_{-}E_{+}e^{i(\phi_{+} - \phi_{-})}e^{i2\omega_{m}t}~+ \\
    \nonumber
    & \quad 2\tx{Re}\left[ E_{+}E_{0}e^{i(\phi_{0} - \phi_{+})} + E_{-}E_{0}e^{i(\phi_{-} - \phi_{0})}\right]\cos{(\omega_{m}t)} ~+\\
    & \quad  2\tx{Im}\left[ E_{+}E_{0}e^{i(\phi_{0} - \phi_{+})} + E_{-}E_{0}e^{i(\phi_{-} - \phi_{0})}\right]\sin{(\omega_{m}t)}.
\end{align}
The coefficients of $\sin{(\omega_{m}t)}$ and $\cos{(\omega_{m}t)}$ are different quadratures, real and imaginary components, of a radio-frequency signal at $\omega_{m}$, and encodes the PDH error signal $\boldsymbol{\epsilon}$ that provides a sensitive measure of the frequency shift induced by the qubit on the cavity. Extracting $\boldsymbol{\epsilon}$ requires a combination of filtration and demodulation using an RF LO at $\omega_{m}$, either purely in analog, or via digitization and digital demodulation.  
\begin{figure} [t]
    \centering
    \includegraphics[width=\textwidth]{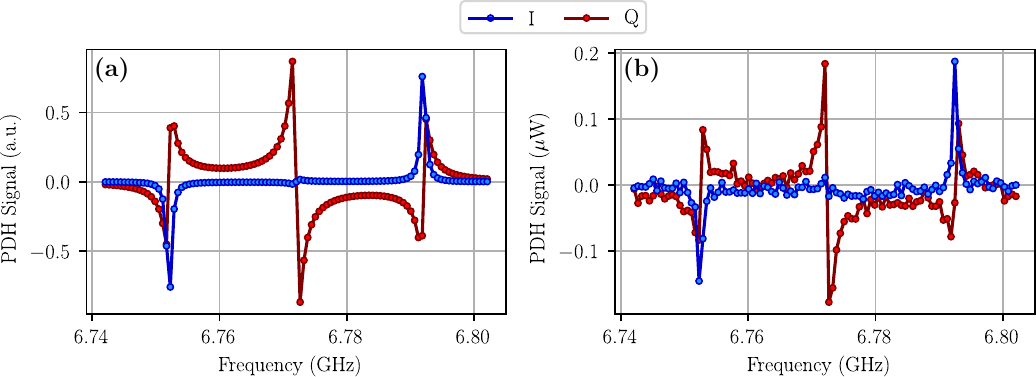}
    \vspace{-0.2cm}
    \caption{\textit{Pound-Drever-Hall (PDH) error signal.} (a) Theoretical PDH error signal, with the real component in blue and the imaginary component in red. (b) Real (I, blue) and imaginary (Q, red) quadratures of the reconstructed PDH error signal obtained with a 20 MHz modulation frequency. The Q component exhibits a steep linear slope near cavity resonance (6.7725~GHz), which is traditionally used for laser locking, and can be used for qubit readout. Data presented here are taken with 1000 averages per point, with the qubit prepared in $\ket{g}$, and without the use of a Josephson parametric amplifier (JPA). (See Appendix \ref{app:deviceandwiring} for description of wiring and JPA.)}
    \label{fig:supple_PDH_ErrorSignal}
\end{figure}
When $\sin{(\omega_{m}t)}$ is used as the RF LO, Im($\boldsymbol{\epsilon}$) is shifted to DC, while all other terms are shifted to $\omega_{m}$ and higher harmonics. These can be removed with a low-pass filter below $\omega_{m}$ to extract Im($\boldsymbol{\epsilon}$). Similarly, when $\cos{(\omega_{m}t)}$ is used, Re($\boldsymbol{\epsilon}$) is extracted. With two RF LOs, or two instances of digital downconversion, the full complex-valued PDH error signal 
\begin{align} \label{eq:pdh_error_signal}
    \boldsymbol{\epsilon}(\omega) &= E_{+}E_{0}e^{i(\phi_{0} - \phi_{+})} + E_{-}E_{0}e^{i(\phi_{-} - \phi_{0})}
\end{align}
can be extracted.

Equations \ref{eq:power_signal2} and \ref{eq:pdh_error_signal} can be written very conveniently in terms of the complex beatnotes between the carrier and the two sidebands
\begin{equation}
    \boldsymbol{B}_{\pm} = E_{\pm}E_{0}e^{\pm i(\phi_{0} - \phi_{\pm})}.
\end{equation}
This form is particularly convenient for analyzing the changes in $\boldsymbol{\epsilon}$ due to imperfections in the measurement setup. In this notation, 
\begin{align}
    \nonumber
    \boldsymbol{\epsilon} (\omega) = \boldsymbol{B}_{+} + \boldsymbol{B}_{-}.
\end{align}
As described above, the error signal has two components or quadratures: 
\begin{align} \label{eq:pdh_I}
    \nonumber
    \epsilon_{\tx{I}}(\omega) &= \tx{Re}(\boldsymbol{\epsilon}) \\
    \nonumber
    & = \tx{Re}(\boldsymbol{B}_{+}) + \tx{Re}(\boldsymbol{B}_{-}) \\
    & = E_{+}E_{0}\cos{(\phi_{0} - \phi_{+})} + E_{-}E_{0}\cos{(\phi_{0} - \phi_{-})},
\end{align}
extracted using $\cos{(\omega_{m}t)}$, and 
\begin{align} \label{eq:pdh_Q}
    \nonumber
    \epsilon_{\tx{Q}}(\omega) &= \tx{Im}(\boldsymbol{\epsilon}) \\
    \nonumber
     & = \tx{Im}(\boldsymbol{B}_{+}) + \tx{Im}(\boldsymbol{B}_{-}) \\
     & = E_{+}E_{0}\sin{(\phi_{0} - \phi_{+})} - E_{-}E_{0}\sin{(\phi_{0} - \phi_{-})},
\end{align}
extracted using $\sin{(\omega_{m}t)}$. Note that the convention used here is different from the standard optics convention where $\pi$ is explicitly factored out from $\bar{E}_{-}e^{i(\omega_{-}t + \bar{\phi}_{-})} \to -\bar{E}_{-}e^{i(\omega_{-}t + \bar{\phi}'_{-})}$, which gives the more conventional equation
\begin{align}
    \nonumber
    \epsilon_{\tx{Q}}(\omega) = E_{+}E_{0}\sin{(\phi_{0} - \phi_{+})} + E_{-}E_{0}\sin{(\phi_{0} - \phi'_{-})}.
\end{align}

The frequency dependence of the two components of the PDH error signal is shown in \fref{fig:supple_PDH_ErrorSignal} with the theoretical simulation (\fref{fig:supple_PDH_ErrorSignal}(a)) calculated with parameters ($\omega_{0} = 6.7725$~GHz, $\omega_{m} = 20$~MHz) specifically chosen to mirror the experimental results in \fref{fig:supple_PDH_ErrorSignal}(b). The measured quadratures in \fref{fig:supple_PDH_ErrorSignal}(b), taken with an average of $1000$ single-shots per frequency point, are not obtained using the PDH scheme described above (due to the absence of a suitable cryogenic power detector), but from a reconstruction of the PDH error signals using the triple downconversion method which will be presented in Appendix \ref{sec:TripleDDC}. The component $\epsilon_{\tx{Q}}$ plays the central role in carrying the qubit-state information. In contrast to $\epsilon_{\tx{I}}$, which ideally vanishes and contributes no usable signal for state discrimination, $\epsilon_{\tx{Q}}$ exhibits distinct dispersive features around resonance that are sensitive to the qubit-state-dependent frequency shift of the readout resonator. The steep linear slope of $\epsilon_{\tx{Q}}$ near resonance, with its extrema occuring at detunings of approximately $\pm \kappa/2$ from the zero crossing, is effective for converting small frequency shifts into measurable signal variations. By engineering the dispersive shift such that $2\chi \approx \kappa$, the ground and excited states can be positioned near the opposite peaks corresponding to the positive and negative extrema of $\epsilon_{\tx{Q}}$ near resonance. Such a configuration maximizes the separation between the two qubit states, enhancing readout and qubit-state discrimination. 

\subsection{Insensitivity of PDH to Phase Errors} \label{app:insensitivity_of_pdh}

The PDH method has proven extremely successful because it allows measurement of the phase response of a cavity without being sensitive to the absolute phases. 
Here, we review the common error channels/sources in a microwave PDH measurement setup and show how $\boldsymbol{\epsilon}$ behaves under typical sources of phase fluctuations.
Crucially, since PDH depends on the beatnotes between the carrier and the sidebands, the effect of experimental imperfections and phase errors on $\boldsymbol{\epsilon}$ are best understood from the perspective of different types of correlated phase shifts effecting all three tones. Overwhelmingly the most common types of errors are either  \emph{common-mode} errors which shift all three tones by the same amount or \emph{differential-mode} phase errors which impart opposite phase shifts to the two sidebands and leave the carrier unchanged.

Under the effect of a common-mode phase error of strength $\alpha$, the three measured phases $\left( \phi_-, \phi_0, \phi_+ \right)$ transform into three new phases
\begin{equation}
 \left( \phi_-', \phi_0', \phi_+' \right) = \left( \phi_-, \phi_0, \phi_+ \right) + \alpha (1,1,1).
\end{equation}
The PDH signal $\boldsymbol{\epsilon}$ becomes a function of the modified phases; however, 
\begin{align}
    \nonumber
    \phi'_{0} - \phi'_{\pm} &= \phi_{0} + \alpha - \phi_{\pm} - \alpha \\
    \nonumber
    &= \phi_{0} - \phi_{\pm}.
\end{align}
Thus, both $\boldsymbol{B}_{-}$ and $\boldsymbol{B}_{+}$ remain unchanged and, by extension, $\boldsymbol{\epsilon}$. In contrast, heterodyne readout, which depends on $\phi_0$ directly, will be phase shifted by $\alpha$.

Differential-mode phase errors do not affect the carrier and therefore have no effect on heterodyne readout. However, they do effect the PDH signal. 
Under the effect of a differential-mode phase error of strength $\beta$, the measured phases of the three PDH tones transform to 
\begin{equation}
 \left( \phi_-', \phi_0', \phi_+' \right) = \left( \phi_-, \phi_0, \phi_+ \right) + \beta (-1,0,1).
\end{equation}
Under these conditions
$$
    \phi'_{0} - \phi'_{\pm} = \phi_{0} - \phi_{\pm} \mp \beta,
$$
\begin{align}
    \nonumber
    \boldsymbol {B}'_{+} &= E_{+}E_{0}e^{i(\phi_{0} - \phi_{+})}e^{-i\beta} \\
    \nonumber
    &= \boldsymbol{B}_{+}e^{-i\beta},
\end{align}
and 
\begin{align}
    \nonumber
    \boldsymbol{B}'_{-} &= E_{-}E_{0}e^{i(\phi_{-} - \phi_{0})}e^{-i\beta} \\
    \nonumber
    &= \boldsymbol{B}_{-}e^{-i\beta}.
\end{align}
As a result of these phase shifts in the beatnotes, $\epsilon_{\tx{I}} \to \epsilon_{\tx{I}}'$, where 
\begin{align} \label{eq:pdh_I_prime}
    \nonumber
    \epsilon_{\tx{I}}' &= \tx{Re}(\boldsymbol{B'_{+}}) - \tx{Re}(\boldsymbol{B'_{-}}) \\
    \nonumber
    & = \tx{Re}(\boldsymbol{B}_{+}e^{-i\beta}) - \tx{Re}(\boldsymbol{B}_{-}e^{-i\beta}) \\
    \nonumber
    & = \left[\tx{Re}(\boldsymbol{B}_{+}) - \tx{Re}(\boldsymbol{B}_{-})\right]\cos{(\beta)} + \left[\tx{Im}(\boldsymbol{B}_{+}) - \tx{Im}(\boldsymbol{B}_{-})\right]\sin{(\beta)} \\
    & = \epsilon_{\tx{I}}\cos{(\beta)} + \epsilon_{\tx{Q}}\sin{(\beta)}.
\end{align}
Similarly, $\epsilon_{\tx{Q}} \to \epsilon_{\tx{Q}}'$, where
\begin{align} \label{eq:pdh_Q_prime}
    \nonumber
    \epsilon_{\tx{Q}}' &= \tx{Im}(\boldsymbol{B'_{+}}) - \tx{Im}(\boldsymbol{B'_{-}}) \\
    \nonumber
    & = \tx{Im}(\boldsymbol{B}_{+}e^{-i\beta}) - \tx{Im}(\boldsymbol{B}_{-}e^{-i\beta}) \\
    \nonumber
    & = \left[\tx{Im}(\boldsymbol{B}_{+}) - \tx{Im}(\boldsymbol{B}_{-})\right]\cos{(\beta)} + \left[-\tx{Re}(\boldsymbol{B}_{+}) + \tx{Re}(\boldsymbol{B}_{+})\right]\sin{(\beta)} \\
    & = \epsilon_{\tx{Q}}\cos{(\beta)} - \epsilon_{\tx{I}}\sin{(\beta)}.
\end{align}
Equations \ref{eq:pdh_I_prime} and \ref{eq:pdh_Q_prime} show that differential-mode phase errors cause a rotation between the quadratures of the PDH error signal. However, as will be discussed below, the hardware conditions required to achieve a large differential-mode phase error are typically extreme.

Next we review the types of phase errors typically produced by the hardware components in a superconducting-qubit readout setup and their effect on $\boldsymbol{\epsilon}$.

\vspace{0.08in}
\noindent \textit{Generator phase errors}: Since all the three tones originate from the same microwave generator (see \fref{fig:wiringDiagram}), any frequency or phase error of the generator is imprinted identically onto each tone, resulting in a purely common-mode phase error, which is intrinsically subtracted out in the PDH error signal. 
As a result, qubit-state readout with PDH still works well even in the presence of large generator phase errors, unlike conventional heterodyne readout under the same conditions.
Single-shot readout data demonstrating this resilience of PDH to generator-induced common-mode phase errors is shown in Fig. 3(a) of the main text. 

\vspace{0.08in}
\noindent \textit{Modulation phase errors}: The modulation that generates the sidebands is created by an RF source, Zurich Instruments HDAWG in this work (see \fref{fig:wiringDiagram}). The phase errors introduced by the RF source affect only the sidebands, producing only a differential-mode phase error and leaving the carrier unaffected. This will rotate the PDH quadratures. However, the modulation phase error resulting from a time delay $\Delta t$ in the RF source is $\beta = \omega_m \Delta t$, and will be insignificant unless the time delay is large compared to the period of the (RF) modulation. 

\vspace{0.08in}
\noindent \textit{Path length/propagation fluctuations}: For a signal propagating along a path of length $L$ at frequency $\omega$ and phase velocity $\upsilon_{p}$, the phase fluctation $\Delta \phi$ caused by a path length fluctuation $\Delta L$ is given by $\Delta \phi = \omega \Delta L/\upsilon_{p}$. Since the three PDH tones propagate along the same path, the resulting phase contributions to each tone from a path length fluctuation are correlated such that 
\begin{align}
    \nonumber
    (\phi'_{-}, \phi'_{0}, \phi'_{+}) & =  \left( \phi_{-} + \frac{\omega_{-}\Delta L}{\upsilon_{p}}, \phi_{0}+\frac{\omega_{0}\Delta L}{\upsilon_{p}}, \phi_{+}+\frac{\omega_{+}\Delta L}{\upsilon_{p}} \right) \\
    \nonumber
    & = (\phi_{-}, \phi_{0}, \phi_{+}) + \alpha(1,1,1) + \beta(-1, 0, 1),
\end{align}
where
\begin{align}
    \nonumber
    \alpha = \frac{\omega_{0}\Delta L}{\upsilon_{p}}
\end{align}
is the common-mode phase fluctuation, and 
\begin{align}
    \nonumber
    \beta = \frac{\omega_{m}\Delta L}{\upsilon_{p}} \ll \alpha
\end{align}
is the differential-mode phase change. Hence, the total effect of a path length fluctuation manifests as a combination of the \emph{common-mode} and \emph{differential-mode} phase errors discussed above. Under these conditions, heterodyne would be phase shifted by $\alpha$. PDH, on the other hand, is only sensitive to the differential-mode phase errors $\beta$, which is much smaller than $\alpha$, and only significant when the path length fluctuation is comparable to the wavelength of the modulation frequency.  

\vspace{0.08in}
\noindent \textit{Digitizer timing errors}: Similarly, phase fluctuations due to the digitizer timing errors also induce both common-mode and differential-mode phase errors such that
\begin{align}
    \nonumber
    (\phi'_{-}, \phi'_{0}, \phi'_{+}) &= (\phi_{-} +\omega_{-}\Delta t, \phi_{0}+\omega_{0}\Delta t, \phi_{+}+\omega_{+}\Delta t) \\
    \nonumber
    & = (\phi_{-}, \phi_{0}, \phi_{+}) + \alpha(1,1,1) + \beta(-1, 0, 1),
\end{align}
where 
\begin{align}
    \nonumber
    \alpha = \omega_{0}\Delta t,
\end{align}
and 
\begin{align}
    \nonumber
    \beta = \omega_{m}\Delta t \ll \alpha.
\end{align}
Under these conditions, heterodyne would be phase shifted by $\alpha$. PDH, on the other hand, is only sensitive to the differential-mode phase errors $\beta$, which is much smaller than $\alpha$, and only significant when timing error is comparable to the period of the modulation.

\vspace{0.08in}
In summary, absolute phase errors due to generator, path-length, and digitizer timing fluctuations all effect conventional heterodyne readout. In contrast, PDH-style readout is immune to the common-mode component of all of these error processes and only sensitive to the differential-mode component. However, the size of the differential-mode phase errors is suppressed by $\tfrac{\omega_m}{\omega_0} \ll 1$. This combined immunity and error gives PDH phenomenal stability and has made it ubiquitous in optical applications. It also allows stable qubit readout which is highly insensitive to timing and clocking errors.

\begin{figure*}[t!]
    \centering
    \includegraphics[width=0.8\textwidth]{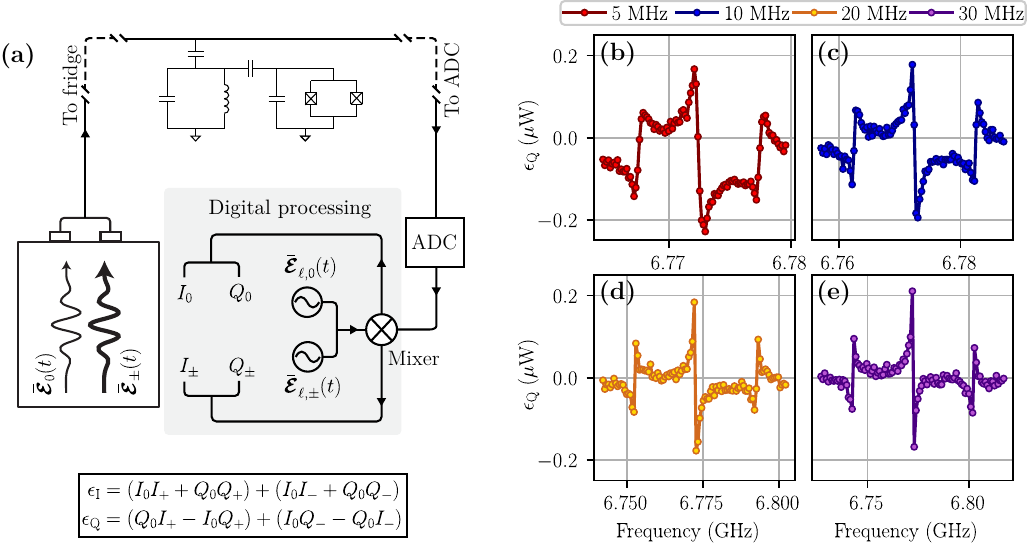}
    \caption{\textit{Synthetic PDH}. (a) Schematic of the triple downconversion setup used to reconstruct the PDH error signal in the absence of a cryogenic power detector. The signal path shown here, from the generator to the ADC, is abbreviated from the full wiring diagram in \fref{fig:wiringDiagram}. The carrier signal $\bar{\boldsymbol{\mathcal{E}}}_{0}(t)$ and the two sidebands $\bar{\boldsymbol{\mathcal{E}}}_{\pm}(t)$ are sent through the dilution refrigerator and simultaneously downconverted by a common analog LO before digitization. Each digitized tone is further digitally demodulated to extract the corresponding IQ quadratures, which are then used to reconstruct the PDH error signal $\boldsymbol{\epsilon}$. (b-e) The reconstructed Q quadrature $\epsilon_{\tx{Q}}$ of the PDH error signal for different modulation frequencies of 5, 10, 20, 30 MHz, obtained from carrier frequency sweeps across the 6.7725 GHz resonance over a span of $3\omega_{m}$. Each frequency scan is taken without the JPA to avoid sideband imbalance from the asymmetric JPA gain profile, and each frequency point is an average of 1000 single shots. (See Appendix \ref{app:deviceandwiring} for details.) 
    }
    \label{fig:supple_readoutSchematics_ModFreqSweep}
\end{figure*}
\subsection{Synthetic PDH without a power detector} \label{sec:TripleDDC}
In the absence of a suitable cryogenic square-law power detector or a microwave equivalent of a photodetector, we implement PDH-style readout by reconstructing $\boldsymbol{\epsilon}$ from the individual tones obtained via simultaneous heterodyne detection of the carrier and the two sidebands, as was done in simultaneous independent work on resonator characterization~\cite{ZurichPDH}. After interacting with the device under test and undergoing amplification, the three PDH tones are mixed with a common LO at frequency $\omega_{\ell}$, resulting in downconverted signals at frequencies ($\omega_{-} - \omega_{\ell}$, $\omega_{0}-\omega_{\ell}$, $\omega_{+}-\omega_{\ell}$) = ($\omega_{\tx{IF}} - \omega_{m}$, $\omega_{\tx{IF}}$, $\omega_{\tx{IF}}+\omega_{m}$). After digital demodulation at $\omega_{\tx{IF}} - \omega_{m}$, $\omega_{\tx{IF}}$, $\omega_{\tx{IF}}+\omega_{m}$ for the lower sideband, the carrier, and the upper sideband, respectively, the corresponding quadratures $[(I_{-}, Q_{-}), (I_{0}, Q_{0}), (I_{+}, Q_{+})]$ are extracted. A schematic for the triple downconversion setup used for the reconstruction is as shown in \fref{fig:supple_readoutSchematics_ModFreqSweep}(a). 

From the three sets of quadratures, we reconstruct the PDH error signal as follows. We formulate a compact cartesian notation for the error signal by introducing a 3D vector $\textbf{v}(\omega) = \left[I(\omega), Q(\omega), 0\right]$ corresponding to each measurement. Using this notation the PDH error signal can be rewritten as
\begin{align} \label{eq:pdh_I_cross-dot}
    \nonumber
    \epsilon_{\tx{I}}(\omega) & = E_{+}E_{0}\cos{(\phi_{0} - \phi_{+})}  + E_{-}E_{0}\cos{(\phi_{0} - \phi_{-})} \\ 
    \nonumber
    & = \left[I_{0}I_{+} + Q_{0}Q_{+}\right] + \left[I_{0}I_{-} + Q_{0}Q_{-}\right] \\
    & = \left[\textbf{v}_{+}(\omega_{+}) \cdot \textbf{v}_{0}(\omega_{0})\right] + \left[\textbf{v}_{0}(\omega_{0}) \cdot \textbf{v}_{-}(\omega_{-})\right].
\end{align}
Similarly,
\begin{align} \label{eq:pdh_Q_cross-dot}
    \nonumber
    \epsilon_{\tx{Q}}(\omega)& = E_{+}E_{0}\sin{(\phi_{0} - \phi_{+})} - E_{-}E_{0}\sin{(\phi_{0} - \phi_{-})} \\ 
    \nonumber
    & = (Q_{0}I_{+} - I_{0}Q_{+}) + (I_{0}Q_{-} -  Q_{0}I_{-}) \\
    & = \left[\textbf{v}_{+}(\omega_{+}) \times \textbf{v}_{0}(\omega_{0})\right]_{\tx{z}} - \left[\textbf{v}_{-}(\omega_{-}) \times \textbf{v}_{0}(\omega_{0})\right]_{\tx{z}}.
\end{align}
Formulating the PDH signal in this way allows $\epsilon_{\tx{I}}$, $\epsilon_{\tx{Q}}$ to be reconstructed without needing to extract $\phi_{0,\pm}$ directly, an extraction which is poorly behaved for noisy signals.

Reconstructed PDH error signals $\epsilon_{\tx{Q}}$ obtained for probe frequency sweeps of $\pm3 \omega_{m}$ across the resonance at $6.7725$~GHz are presented in \fref{fig:supple_readoutSchematics_ModFreqSweep}(b-e). At each probe frequency, $1000$ single shots are averaged to obtain a single IQ point for each tone (lower sideband, carrier, upper sideband), and used to compute the reconstructed PDH signal using \eqsandref{eq:pdh_I_cross-dot}{eq:pdh_Q_cross-dot}, for modulation frequencies $\omega_{m}/2\pi$ = $ \{5, 10, 20, 30\}$~MHz. To avoid sideband imbalance from the asymmetric JPA gain profile across the sweep range (see Appendix \ref{app:deviceandwiring} for details), these measurements (and other frequency-sweep measurments reported here) were performed without parametric amplification (\fref{fig:wiringDiagram}).

\begin{figure*}[t]
    \centering
    \includegraphics[width=0.9\textwidth]{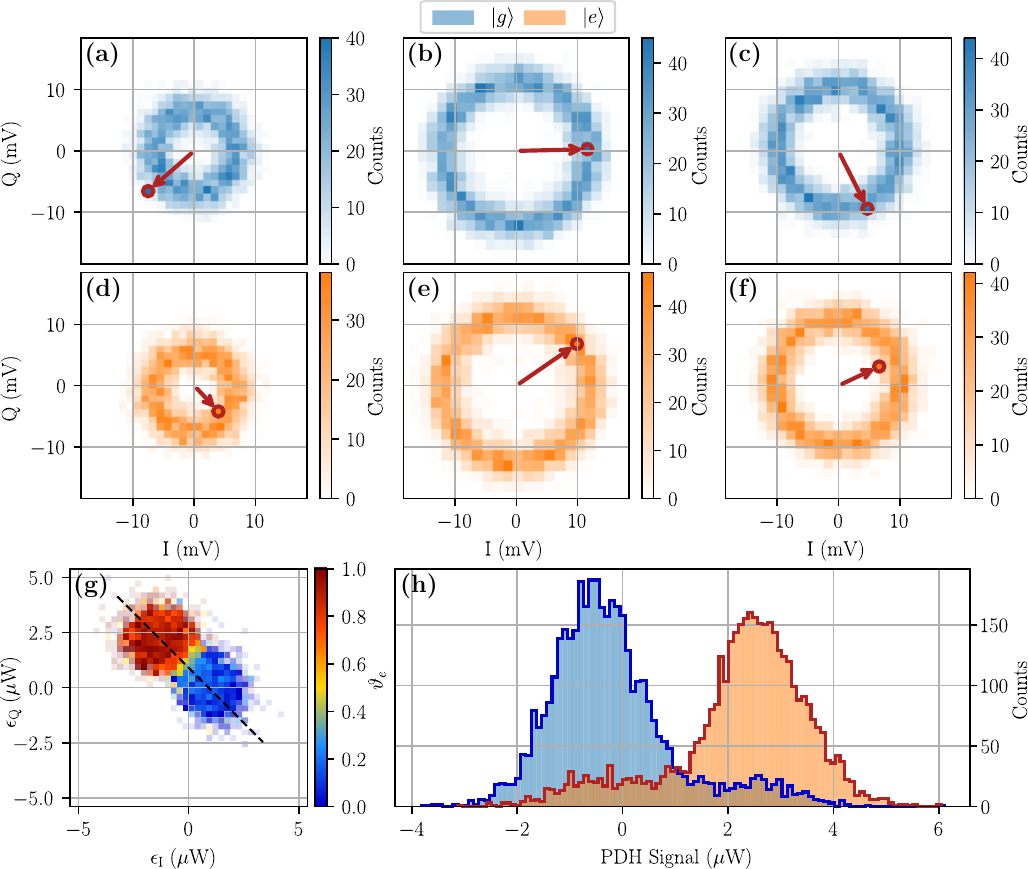}
    \caption{\textit{Robustness of PDH to common-mode phase errors.} 5000 Single-shot IQ measurements are taken with the carrier generator unlocked from the external reference, both when the qubit is prepared in $\ket{g}$(blue) and when it is prepared in $\ket{e}$(orange). (a-c) IQ distribution of the lower sideband, carrier, and upper sideband respectively, when the qubit is prepared in $\ket{g}$. Single-shot outcomes of the $50$th acquisition are indicated by the red arrows. The induced common-mode phase errors on the three tones compromise the absolute phase stability; however, the three tones remain coherent with eachother. (d-f) Same measurements as in (a-c) but with the qubit prepared in the excited state. (g) Reconstructed PDH IQ components from the three tones. The color bar illustrates $\vartheta_{e}$, the fraction of measurements per pixel corresponding to preparation in $\ket{e}$. Bins with $\vartheta_{e}=1$ (red) correspond to IQ locations originating purely from $\ket{e}$, $\vartheta_{e}=0$ (blue) to locations originating purely from $\ket{g}$, and $\vartheta_{e}=0.5$ (gold) to IQ locations for which both preparations are equally likely.  For bins with less than 10 counts, transparency is scaled with the number of counts, since $\vartheta_{e}$ cannot be accurately computed for these pixels. This ensures that only statistically meaningful regions contribute to the structure of the plotted distribution and noise-dominated regions are not visually overemphasized. The black dashed line denotes the axis of maximum state separation, which deviates from the Q axis due to phase shifts between the modulation and demodulation, and gain asymmetry of the JPA (\fref{fig:wiringDiagram}). (h) Histogram of the final projected PDH signal along the axis of maximum separation. Clear separation between typical ground-state and excited-state outcomes is visible despite the lack of phase coherence in the individual tones. Residual overlap between the two distributions is primarily due to state preparation errors or decay during measurement.}
    \label{fig:supple_carrierUnlocked_singleShot}
\end{figure*}
\begin{figure*}[t]
    \centering
    \includegraphics[width=0.48\textwidth]{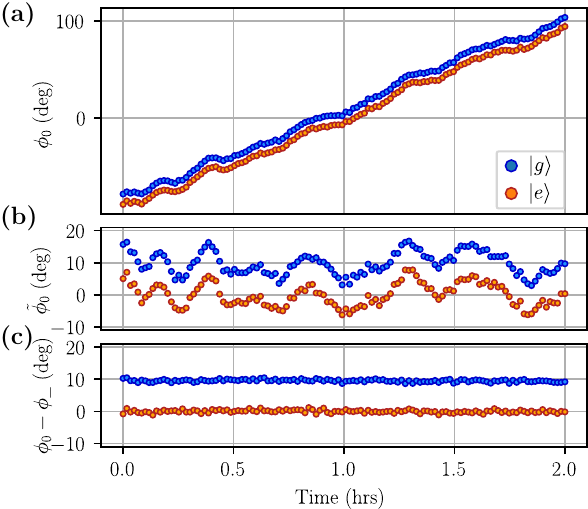}
    \caption{\textit{Long-term phase stability of qubit readout.} Average carrier phase from $1000$ consecutive single shots of the qubit prepared in $\ket{g}$ (blue) and in $\ket{e}$(orange), measured intermitently over a period of two hours. The mean phase of the $\ket{e}$ measurements is subtracted from all traces. (a) Raw carrier phase $\phi_{0}$ showing a linear drift of ~200$^\circ$ for both $\ket{g}$ and $\ket{e}$, corresponding to a frequency offset of ~73$\mu$Hz between the carrier and LO generators. (b) Residual non-linear carrier phase $\tilde{\phi}_{0}$ for the qubit states after removing the linear drift in (a). (c) Differential phase $\phi_{0} - \phi_{-}$ from the PDH-style readout, showing stable and well-separated qubit states over the same time period.}
    \label{fig:supple_phaseStability_2hrs}
\end{figure*}
\section{Phase Stability}\label{sec:phasestability}
To evaluate the robustness of PDH readout against phase and timing errors experimentally, we carry out single-shot measurements of the qubit state via reconstructed PDH while deliberately removing the phase-lock between the carrier generator and the external reference clock. 
With the carrier generator unlocked, we measure $5000$ single-shots while preparing the qubit in either the ground or excited state and recording the IQ components of all 3 PDH tones. The phase of each individual tone becomes random due to the loss of stable phase reference. The resulting IQ distributions of the three tones, when the qubit is prepared in either the ground state or the excited state, are shown in  \fref{fig:supple_carrierUnlocked_singleShot}(a-c) and \fref{fig:supple_carrierUnlocked_singleShot}(d-f), respectively. The IQ distributions for the two qubit states are nearly indistinguishable (see Fig~3(a) in main text) making good state discrimination impossible. In contrast, the single-shot PDH signal reconstructed from these tones,  remains phase stable due to carrier-sideband phase coherence and yields two distinct single-shot clusters corresponding to the ground and excited states of the qubit (\fref{fig:supple_carrierUnlocked_singleShot}(g)). 

Single-shot IQ distributions corresponding to $\ket{g}$ and $\ket{e}$ state preparation are shown separately in \fref{fig:supple_carrierUnlocked_singleShot}(a-f). In Fig.~3(a) of the main text and \fref{fig:supple_carrierUnlocked_singleShot}(g), we present the heterodyne and PDH outcomes corresponding to both states together, and examine the relative likelihood of $\ket{g}$ and $\ket{e}$ outcomes across
the IQ plane to highlight the regions of overlap or separation between the two states. For this analysis, we use $5000$ single-shot measurements for each state. The IQ plane of the combined IQ measurements of the two states is binned into $40 \times 40$ uniform bins. 
For each bin $(x,y)$, we compute the fraction of measurements corresponding to preparation in $\ket{e}$
\begin{align}
    \nonumber
    \vartheta_{e}(x,y) = \frac{n_{e}(x,y)}{n_{tot}(x,y)},
\end{align}
where $n_{e}$ and $n_{tot}$ are the instances of $\ket{e}$ outcomes and the total number of outcomes in that bin, respectively. Colormaps of $\vartheta_{e}$ for the heterodyne and PDH readout with free running generators are shown in Fig.~3(a) in the main text and \fref{fig:supple_carrierUnlocked_singleShot}(g) respectively.
In bins with very few measurement outcomes, $\vartheta_{e}$ is likely dominated by shot noise and not a faithful indicator of overlap between $\ket{g}$ and $\ket{e}$ outcomes. To prevent such pixels from being visually overemphasized, we reduce the transparency of bins with $n_{tot}<10$. This ensures that only statistically meaningful regions contribute to appearance of the distribution, thus more accurately reflecting the underlying data distribution. 

The two-dimensional PDH IQ distributions can be simplified by projection onto the axis that maximizes the distance between the ground state and the excited state outcomes. For an ideal PDH configuration, this optimal axis of maximum separation aligns with the Q quadrature because the I quadrature carries no signal contrast. However, phase shifts between the modulation and demodulation generally rotate the measurement axis away from the nominal Q direction. In this work, possible line dispersion in the readout path and unequal amplification of the two sidebands due to an asymmetric JPA gain profile  ($E_{+} \neq E_{-}$) provide additional shifts. (See Appendix \ref{app:deviceandwiring} for discussion of JPA and its offsets.) To account for the imbalance causing such rotation, we consider two types of corrections: (i) post-processing corrections (used here) done to recover the correct readout axis and (ii) pre-processing corrections (described in the next Appendix) which involve precompensating the generated phases of the three tones to counter the expected phase acquisition of the sidebands along the readout path. For the case here where the readout information is in both quadratures, we compute the centroids of each distribution and define a projection axis as the normalized vector pointing from the ground to the excited state centroid (black dashed line in \fref{fig:supple_carrierUnlocked_singleShot}(g)). Each single-shot IQ point is then projected onto this axis, resulting in a one-dimensional distribution that represents the effective final PDH readout signal. 

The single-shot measurements shown in \fref{fig:supple_carrierUnlocked_singleShot} demonstrate that PDH readout remains robust against common-mode errors on short timescales. To evaluate its long-term stability under optimal lab conditions, where all signal sources are locked to an external reference clock, we monitor the readout phase for two hours with the qubit prepared alternatively in $\ket{g}$ and $\ket{e}$ and with a $30~$MHz sideband applied (\fref{fig:supple_phaseStability_2hrs}). During the two-hour measurement, we compute the average phase of 1000 consecutive single shots every minute. For conventional heterodyne readout, the averaged carrier phase $\phi_{0}$ for both qubit states exhibits a linear drift of roughly $200^{\circ}$ corresponding to a frequency offset of 73~$\mu$Hz between the main and LO generators, primarily due to mismatched generators models. The measured raw carrier phase $\phi_{0}$ for $\ket{g}$ and $\ket{e}$ remain distinct at any given time but the drift far exceeds the readout signal, as shown in \fref{fig:supple_phaseStability_2hrs}(a). Even after removing the linear drift in post processing, the resulting non-linear drift $\tilde{\phi}_{0}$ is also comparable to the separation between the two qubit states (\fref{fig:supple_phaseStability_2hrs}(b)). In contrast, the differential phase $(\phi_{0} - \phi_{-})$ relevant to PDH readout remains extremely stable with average RMS of $0.44^{\circ}$ over the entire two-hour period, as shown in \fref{fig:supple_phaseStability_2hrs}(c). This indicates that fluctuations in laboratory conditions or clocking of the measurement components are not producing noticeable differential-mode phase errors.
Measurements corresponding to the two qubit states remain clearly separated and consistent at all times, establishing the exceptional phase stability of PDH readout.

\begin{figure*} [t!] 
    \centering
    \includegraphics[width=0.9\textwidth]{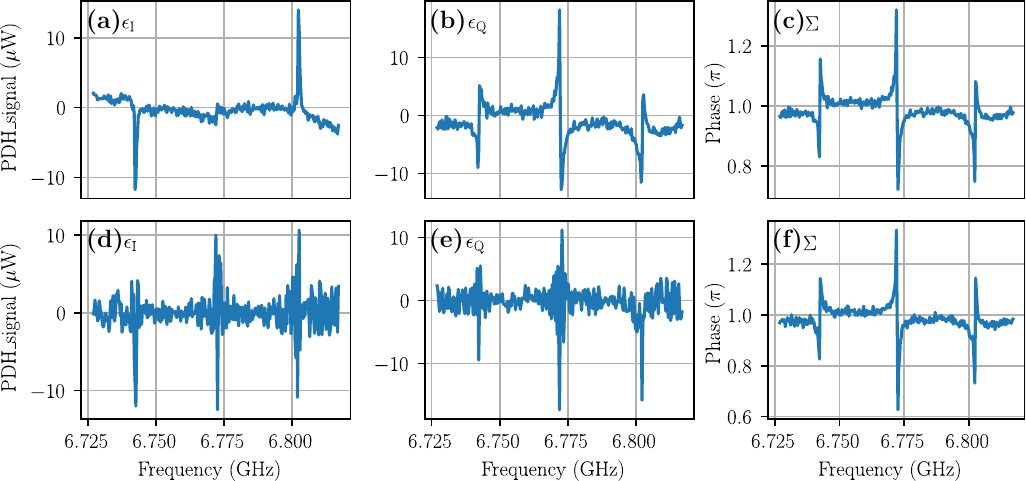}
    \vspace{-0.2cm}
    \caption{\textit{Robustness of scissors phase $\Sigma$ to timing errors.} (a-c)  $\epsilon_{\tx{I}}$, $\epsilon_{\tx{Q}}$, and scissors phase $\Sigma$, respectively, acquired in ideal conditions without any induced timing errors. (d-f) Same quantities as (a-c) except with intentionally induced digitizer and AWG timing errors which produce both common-mode and differential-mode phase errors. The differential-mode error mixes the two quadratures of the PDH error signal, distorting $\epsilon_{\tx{I}}$ and  $\epsilon_{\tx{Q}}$, while $\Sigma$ remains unchanged, demonstrating its robustness to \emph{both} common-mode and differential-mode phase errors.}
    \label{fig:TimingErrorSweep_PDHvSigma}
\end{figure*}

\section{Scissors Phase} \label{sec:scissors_phase}
While PDH readout is inherently robust to common-mode errors and path-length fluctuations, it is still sensitive to differential-mode errors. However, a closely related quantity
\begin{align}
    \nonumber
    \Sigma = 2\phi_{0} - (\phi_{-}+\phi_{+}) = \left( \phi_{-}, \phi_{0}, \phi_{+} \right) \cdot (-1,2,-1)
\end{align}
can be defined, which is immune to both of these types of errors. We refer to this quantity as the scissors phase, in analogy to the normal modes of ion traps \cite{exploringTheQuantum}.

Here, we show that $\Sigma$ is a more general and intrinsically stable indicator of the cavity response than the traditional PDH error signal itself. Normal errors, which are either common-mode or differential, such as path-length fluctuations, timing errors, and drifts in the modulation phases transform the phases from $(\phi_{-}, \phi_{0},\phi_{+})$ to $(\phi'_{-}, \phi'_{0},\phi'_{+})$ where 
\begin{align}
    \nonumber
    (\phi'_{-}, \phi'_{0},\phi'_{+}) = (\phi_{-}, \phi_{0},\phi_{+}) + \alpha(1,1,1) + \beta(-1,0,1).
\end{align}
After such a fluctuation, the transformed scissors phase $\Sigma'$ becomes
\begin{align}
    \nonumber
    \Sigma' & = 2\phi'_{0} - (\phi'_{-}+\phi'_{+}) \\
    \nonumber
    & =  2\phi_{0} - (\phi_{-}+\phi_{+}) \\
    \nonumber 
    & = \Sigma.
\end{align}
Thus, the scissors phase $\Sigma$ is orthogonal to normal errors. However, it is still sensitive to the detuning between the carrier and the cavity, and can be used for qubit readout even in regimes dominated by timing or differential-mode errors. In general, the PDH error signal can be viewed as an approximation to $\Sigma$. In the weak coupling limit $|\phi_{0}-\phi_{\pm}| \ll 1$, $\epsilon_{\tx{Q}}$ can be approximated as 
\begin{align}
    \epsilon_{\tx{Q}} \approx E_{-}E_{0}(\phi_{0}-\phi_{-}) + E_{+}E_{0}(\phi_{0}-\phi_{+}).  
\end{align}
Therefore, if the sideband amplitudes are perfectly balanced, and if the RF LO phase is set to exactly extract $\epsilon_{\tx{Q}}$, then the measured PDH signal will be proportional to $\Sigma$. However, any sideband imbalance or differential-mode phase error will cause the experimentally measured value to deviate from $\Sigma$. Since we are not using a photodiode to directly obtain the PDH signal, but instead using simultaneous heterodyne detection of all three tones, it is possible to compute $\Sigma$ directly from the phases of heterodyne measurements, rather than $\epsilon_{\tx{Q}}$.

The robustness of $\Sigma$ is demonstrated experimentally by comparing reconstructed phase data taken under ideal synchronized conditions with data acquired while intentionally introducing timing errors. Measurements of PDH signal components and $\Sigma$ as a function of frequency, with and without induced timing errors, and with $\omega_{m}/2\pi = 30$~MHz are shown in \fref{fig:TimingErrorSweep_PDHvSigma}. Each data point represents the average of $1000$ single shots and timing is held constant during all 1000 shots per point. Between different frequency points, a random time delay, introduced to emulate modulation and digitization drifts and comparable to the modulation period, is applied to both the AWG and the digitizer. For the AWG, the delay is randomly chosen up to $32$~ns, while for the digitizer, the delay is randomly chosen up to $224$~ns. Under ideal operation, the traditional PDH components, $\epsilon_{\tx{I}}$ and $\epsilon_{\tx{Q}}$, and the scissors phase $\Sigma$ all reproduce the expected resonant features of the cavity response. However, when deliberate RF phase errors such as the digitizer and modulation timing offsets are applied, $\epsilon_{\tx{I}}$ and $\epsilon_{\tx{Q}}$ exhibit the predicted distortions and quadrature rotations associated with differential mode errors (\fref{fig:TimingErrorSweep_PDHvSigma}(d-e)), while $\Sigma$ remains unchanged from its error-free form as shown in \fref{fig:TimingErrorSweep_PDHvSigma}(f).

To ensure accurate phase reconstruction at the detector, the data shown here were taken with precompensation to ensure that the two sidebands have equal and opposite phase to the carrier at the detector rather than at generation. To do this, an initial measurement was taken to determine the average dispersion imparted on the two sidebands. A fixed precompensation, independent of frequency, was applied to each of the sideband tones to enforce the proper relationship at the detector.

\begin{figure*}[t]
    \centering
    \includegraphics[width=0.8\textwidth]{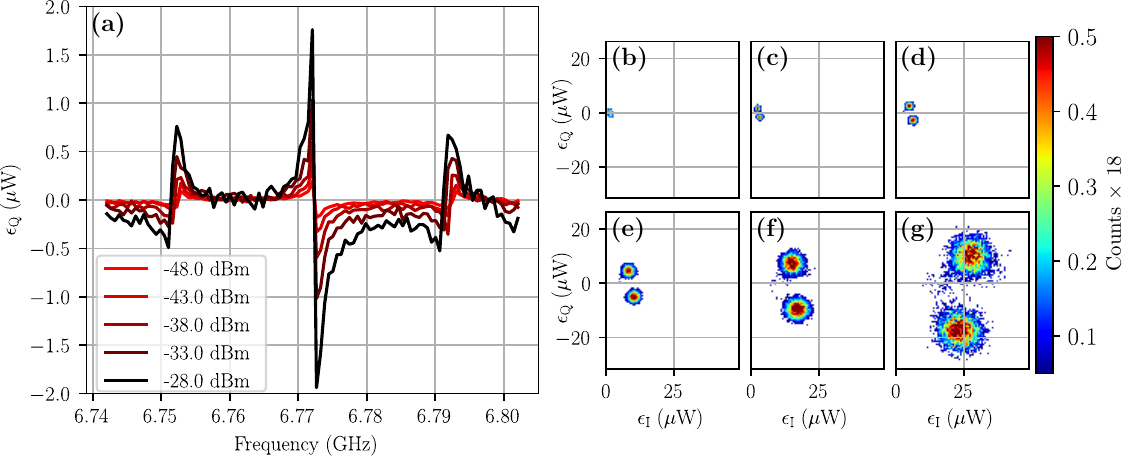}
    \caption{\textit{Reconstruction of heterodyne gain in PDH readout} (a) Reconstructed PDH error signal at a modulation frequency of $\omega_{m}/2\pi = 20$~MHz, obtained while increasing the sideband power from $-48$~dBm to $-28$~dBm with the carrier power fixed at $-45$~dBm. Each measured point is the average of $1000$ single-shot measurements. The amplitude of the PDH response increases with sideband power, consistent with the expected scaling. (b-g) Reconstructed single-shot PDH readout of the two qubit states acquired at $\omega_{m}/2\pi = 15$~MHz for sideband powers ranging from $-50$~dBm to $-25$~dBm and a fixed carrier of $-45$~dBm. While the signal strength grows with increasing sideband power as expected, the overall signal-to-noise ratio remains unchanged because the reconstruction occurs after heterodyne detection of each tone, and therefore inherits all the noise associated with conventional heterodyne readout. The observed scaling in the subfigures illustrates the heterodyne gain expected from PDH, but does not replicate the favorable noise properties expected from a true PDH implementation using a square-law detector.
    }
    \label{fig:PDH_Powerscans}
\end{figure*}

\section{Intrinsic Heterodyne Gain in PDH}
Beyond its phase-stability advantages, the PDH framework also possesses intrinsic heterodyne gain that can enhance readout signal strength compared to heterodyne readout with the carrier alone. This heterodyne gain arises because the PDH readout signal (\eqref{eq:pdh_error_signal}) scales with the product of the carrier and sideband fields. Therefore, when the modulation frequency is large enough to avoid influencing MIST (see Appendix \ref{sec:MIST}), the sideband power can be increased to boost the detected signal without altering the qubit dynamics. However, the measurements presented here are synthetic PDH reconstructed from heterodyne measurements of the individual tones and, therefore, do not directly realize this intrinsic heterodyne gain, since for our setup, all amplification occurs before multiplication of the carruer and sidebands. Below, we simulate the expected scaling behavior for a PDH implementation in which a cryogenic square-law detector is used to mix the carrier and the sidebands \emph{before} amplification.  

To illustrate this scaling, we vary the sideband power from $-48$~dBm to $-28$~dBm while holding the carrier power fixed at $-45$~dBm and reconstruct the average PDH response from the heterodyne data of all three tones with $\omega_{m}/2\pi = 20$~MHz. The overall signal amplitude increases with sideband power, as shown in \fref{fig:PDH_Powerscans}(a). We also examine PDH qubit readout as a function of sideband power by performing 5000 single-shot measurements when the qubit is prepared in $\ket{g}$ and when prepared in $\ket{e}$, as shown in \fref{fig:PDH_Powerscans}(b-g). These measurements were taken with fixed carrier power of $-45$~dBm and $\omega_{m}/2\pi = 15$~MHz, a detuning for which we have demonstrated that high sideband powers do not induce significant added MIST (see \fref{fig:transitionprob_full}). As the sideband power is increased from $-50$~dBm, in \fref{fig:PDH_Powerscans}(b), to $-25$~dBm, in \fref{fig:PDH_Powerscans}(g), the separation between the IQ distributions of the ground and excited states also increases, as expected. Despite the increase in IQ-plance separation with increasing sideband powers, there is no improvement in SNR over the heterodyne readout because for synthetic PDH, all amplifier noise is injected before multiplication. An actual improvement in SNR will require a cryogenic square-law detector,
so that the intrinsic heterodyne gain of PDH is directly synthesized in hardware and takes place before amplification.


\section{State-Assignment}\label{sec:stateassignment}

\begin{figure*}[ht!]
\centering
    \includegraphics[width=\textwidth]{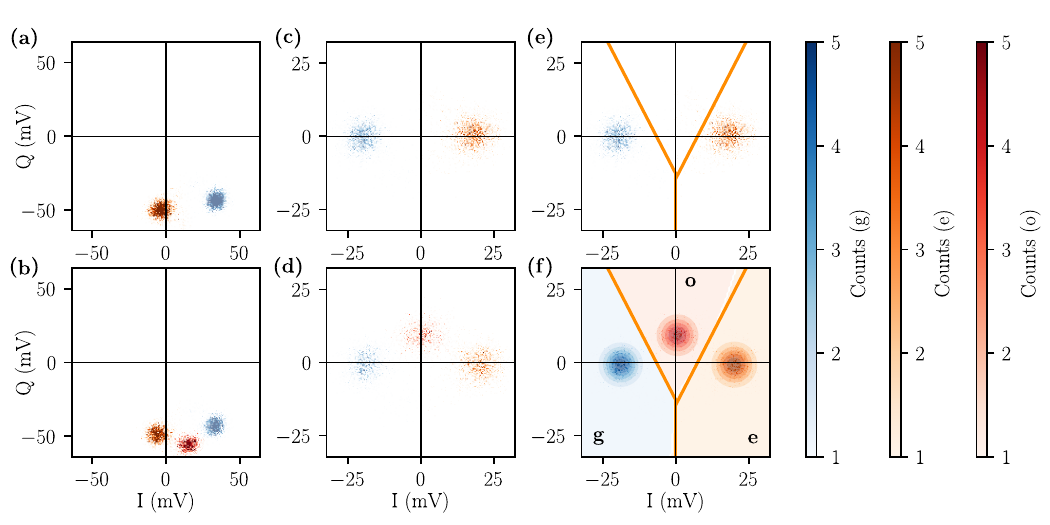}
    \vspace{-0.2cm}
    \caption{\label{fig:fidelitycalc_algorithm} 
    \textit{State-Assignment and Decision Boundary Calculation.}
    Post-analysis of the reference probe configuration, intentionally inducing non-computational states, is performed to obtain the decision boundaries and carry out state assignment. Each column corresponds to a post-processing stage, and each row displays the first (top) and second (bottom) measurements. (a) Raw single-shot IQ histograms.  
    The first measurement in the top row shows two clusters in the IQ plane, corresponding to $\ket{g}$ (blue) and $\ket{e}$ (orange). 
    The second measurement in the bottom row, after the probe is applied, acquires a third cluster, corresponding to $\ket{o}$ (red). Both measurements are analyzed using the \textit{K}-means algorithm to determine the most probable number of clusters and to identify the center of each cluster. 
    The qubit state corresponding to each cluster is inferred from their relative positions.  
    (b) Single-shot IQ histograms after rotation and translation into the standard reference frame, which removes the effect of phase drift and brings all three qubit states near the origin.  
    (c) Decision boundaries (orange lines) obtained from Gaussian fits. 
    Contour plots superimposed on the data show the Gaussian fits for each state. 
    IQ-plane regions corresponding to each qubit state are are shaded in the corresponding color.
    }  
\end{figure*}

Figure~\figqnd~in the main text presents characterization of MIST (measurement-induced state transitions) \cite{Blais:ionization, Blais:ionizationdynamics} experienced by a transmon qubit due to application of a probe tone with variable frequency and power. These MIST characterizations, described in detail in Appendix~\ref{sec:MIST}, benchmark the potential heterodyne gain performance of PDH readout with a cryogenic $|E^2|$ detector. In this Appendix, we describe the state-assignment procedure used to infer the state of the qubit before and after application of the probe tone. The conditional probabilities $p(i|j)$ derived from these assignments are quantitative indicators of harmful MIST and non-QND effects during measurement. 

In conventional superconducting-qubit readout, the measurement is configured to optimize readout fidelity between the two computational-basis states. However, the dominant MIST process typically observed in transmons is transitions to non-computational states~\cite{YaleMIST_fullCharacterization, YaleMIST_highFrequency, google:MISTbeyondRWA}. It is therefore necessary to configure the measurement to provide high visibility between both computational and non-computational states. 
Due to dispersive coupling between the transmon and the readout resonator, each transmon state $\ket{n}$ induces a unique resonator frequency $\omega_{r,\ket{n}}$~\cite{Blais:CavityQED, wallraff2005dispersiveReadout}. 
These unique resonator frequencies impart unique amplitude and phase shifts on the measurement tone, resulting in a different IQ-plane response for each qubit state. Since, $\omega_{r,\ket{n}}$ decreases monotonically with increasing transmon excitation, a measurement frequency of $2\pi \times 6.7717$~GHz near $\omega_{r,\ket{e}}$ and a measurement duration of $2~\mu$s provides optimal distinguishability between the two computational states ($\ket{g}$ and $\ket{e}$) and all "other" higher states, which we denote collectively as $\ket{o}$. Sample IQ-plane outcomes for all three qubit states and this carrier frequency are shown in Fig.~\ref{fig:fidelitycalc_algorithm}.

Determining decision boundaries in the IQ plane which classify single-shot outcomes as either $\ket{g}$, $\ket{e}$, or $\ket{o}$ requires a reference data set in which all three transmon states are present.
Rather than using an independent calibration, we use the MIST characterization scheme itself (see Appendix~\ref{sec:MIST} for details) to obtain all three outcomes,  using a resonant probe of $-38$~dBm 
In this configuration, the first measurement, prior to application of the probe, shows only two possible outcomes, corresponding to the computational states. However, after application of the probe, a third outcome, corresponding to the $\ket{o}$ state appears. These post-probe outcomes are used to determine IQ plane decision boundaries and state assignments for all probe powers and frequencies. Histograms of $5,000$ single-shot IQ measurements in this reference configuration are shown in Fig.~\ref{fig:fidelitycalc_algorithm}(a-b). 

For simplicity, all measurement outcomes are rotated and translated into a standardized coordinate frame, as shown in Fig.~\ref{fig:fidelitycalc_algorithm}(c-d). Since heterodyne measurement is prone to phase drifts in the IQ plane over long times, arising from LO frequency mismatch or residual line dispersion, the reference probe configuration described above is also used as a rolling phase reference. Every probe detuning, a new set of reference IQ histograms is acquired, and the new locations of the three measurement outcomes are then identified, using the procedure described below. The new rotation-translation into the standard reference frame is determined from each reference data set, and updated decision boundaries and state labels are computed, which are then applied to all probe powers at a given detuning.

The rotation-translation into the standard reference frame is determined by applying the \textit{K}-means algorithm~\cite{kmeans_belllab} to the measurement outcomes prior to application of the probe, shown in Fig.~\ref{fig:fidelitycalc_algorithm}(a). The $\ket{g}$ and $\ket{e}$ states are identified by their relative position in the IQ plane, and the center coordinates of the two clusters are used to determine the rotation-translation which brings both computational states to the $x$ axis, symmetrically about the origin. This transformation, shown in Fig.~\ref{fig:fidelitycalc_algorithm}(c), brings the $\ket{o}$ of the transmon onto the positive $y$ axis.

Determining good decision boundaries to discriminate the qubit states requires knowledge of the shapes of the clusters in the measurement outcomes, in addition to their locations. While \textit{K}-means does provide a state assignment for each individual measurement outcome, we find that, for the data presented here, this assignment is prone to systematic errors for events that are not close to the cluster centers. 
Our measurement outcomes are dominated by Gaussian noise originating from photon shot noise and amplifier noise, rather than by qubit lifetime effects or phase drift ~\cite{gambetta2007protocols}. 
Therefore we utilize a Gaussian fit to the reference configuration to determine the decision boundaries and state assignments, rather than the \textit{K}-means clusters directly. 

Under this assumption, the probability distribution of assigning the outcome to the state $\ket{m}$, given a measured value $x$ in 1D or $x,y$ in 2D, is expressed as
\begin{align}\label{eq:gaussian}
    \nonumber
    P_{m} & = \frac{1}{\sigma_m \sqrt{2\pi}} e^{-\frac{(x-m_0)^2}{2\sigma_m^2}} \\
    & = \frac{1}{\sigma_{m_{x}}\sigma_{m_{y}} 2\pi} e^{-\frac{(x-m_{0_{x}})^2}{2\sigma_{m_{x}}^2}} e^{-\frac{(y-m_{0_{y}})^2}{2\sigma_{m_{y}}^2}},
\end{align}
where $m_0$ denotes the center and $\sigma_m$ the variance. The state-assignment fidelity $F_{mn}$ between states $\ket{m}$ and $\ket{n}$, is obtained from the threshold value $x_{th}$ that separates the states:
\begin{align}\label{eq:fidelity}
    \nonumber
    F_{mn} &\equiv 1 - \int_{x_{th}}^{\infty} P_m(x) \, dx - \int_{-\infty}^{x_{th}} P_n(x) \, dx, \\
    &\equiv 1 - \iint_{\mathbb{R}_{n}} P_m(x,y) \, dx dy - \iint_{\mathbb{R}_{m}} P_n(x,y) \, dx dy,
\end{align}
for the one-dimensional and two-dimensional cases, respectively. $\mathbb{R}_{n}$ denotes the region in the IQ plane within which the qubit is assigned to the state $\ket{n}$, defined by the threshold or the decision boundary. For simplicity, we take a simple linear decision boundary between any two states. In general, sophisticated decision boundaries may be necessary to truly maximize single-shot readout fidelity \cite{IBMSVM}. However, since the we are primarily interested in \emph{added} MIST induced by the probe, absolutely optimal measurement parameters and decision boundaries are not necessary.

The parameters of the Gaussian distribution corresponding to each qubit state are determined from a fitting to the second-measurement outcomes of the reference configuration. Initial conditions for the fit are determined by applying \textit{K}-means, and automated confirmation of the presence of three distinct qubit states is carried out using silhouette analysis~\cite{silhouette}. Because the relative positions of the three measurement outcomes in the standard reference frame do not change, each Gaussian cluster can be consistently assigned to a qubit state ($\ket{g}$, $\ket{e}$, or $\ket{o}$).

Using the \textit{K}-means classification and preliminary cluster assignments, the individual qubit-state outcomes are separated and compared pairwise in order to determine the optimal decision boundary. Each pair of clusters is fit to two independent Gaussian distributions to determine the optimal separation line between them. The three pairwise boundaries, orange lines shown in Fig.~\ref{fig:fidelitycalc_algorithm}(f), divide the IQ plane into three disjoint regions, and all single-shot measurement outcomes that fall within these regions are identified as the corresponding qubit states. 

\section{Measurement-Induced State Transitions}\label{sec:MIST}

\begin{figure*}[t]
\centering
    \includegraphics[width=\textwidth]{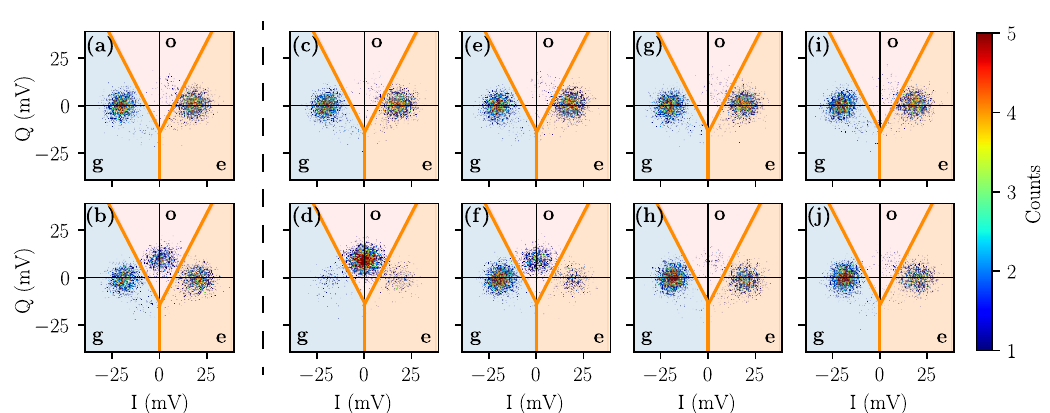}
    \vspace{-0.2cm}
    \caption{\label{fig:detuningsweep_full} 
    \textit{Detuning Dependence of Probe-Induced MIST.}
    IQ histograms of first measurements (top row) and second measurements (bottom row) for $5,000$ single shots versus probe detuning, illustrating the rapid fall-off of probe-induced MIST with detuning.
    (a-b) Reference configuration data used to determine the decision boundaries and qubit-state identification for the probe settings in (c-j). Decision boundaries are indicated in orange, and the IQ-plane regions identified as $\ket{g}$, $\ket{e}$, and $\ket{o}$ are shaded blue, orange, and red, respectively.
    (c-d) At $\Delta_p = 0$, a probe of $-30$~dBm ionizes the transmon almost completely, leaving predominantly $\ket{o}$ outcomes in the second measurement.
    (e-j) Measurement outcomes for the same probe strength as (c-d), but with detunings of $\Delta_p/2\pi = 1,2,20$~MHz, respectively. The $\ket{o}$ population decreases rapidly with detuning, indicating strong suppression of MIST. 
    The residual imbalance between the $\ket{g}$ and $\ket{e}$ populations in (h) and (j) is the result of qubit decay in between the two measurements and is not indicative of probe-induced MIST between the computational basis states.
    }  
\end{figure*}

\begin{figure*}[ht!]
\centering
    \includegraphics[width=\textwidth]{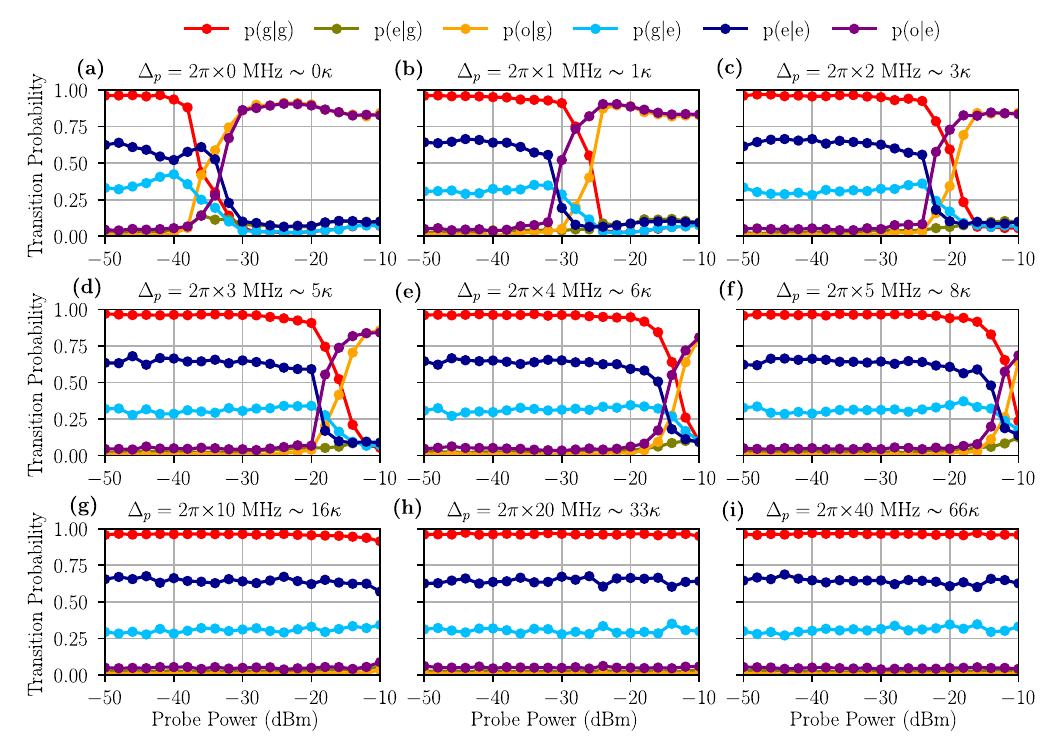}
    \vspace{-0.2cm}
    \caption{\label{fig:transitionprob_full} 
    \textit{Power and Detuning Dependence of Probe-Induced MIST.}
    Power dependence of the transition probabilities $p(i|j)$ versus probe detuning $\Delta_p$, the detuning from the carrier tone. The absolute probe detuning and the detuning with respect to the resonator linewidth $\kappa$ are indicated above each panel. 
    (a) $\Delta_p = 0$. Above a threshold power of approximately $-38$~dBm, a rapid increase in $p(o|g)$ and $p(o|e)$ is observed, accompanied by a simultaneous decrease in the other conditional probabilities, indicating the onset of probe-induced MIST and the breakdown of any readout scheme involving a probe tone of this detuning and power.
    (b–f) For intermediate detunings, upto $\Delta_p \leq 2 \pi\times 5\ \rm{MHz} \sim 8 \kappa$, the breakdown transition increases with frequency, but remains clearly visible within the available dynamic range of probe powers.
    (g) $\Delta_p = 2 \pi\times 10\ \rm{MHz} \sim 16 \kappa$, preliminary signs of breakdown are visible at the highest applied power.
    (h,i) In contrast, for $\Delta_p \geq 2 \pi\times 20 \ \rm{MHz} \sim 33 \kappa$, all transition probabilities remain stable, demonstrating that probe-induced MIST is effectively suppressed for all available probe powers, and that the qubit can tolerate a PDH sideband of at least $+28$~dBc without adverse effects. 
    }  
\end{figure*}

Because the PDH readout signal is proportional to the side-band amplitudes $E_\pm$, large off-resonant sidebands can be used to amplify the total readout signal, leading to heterodyne gain. 
As a result, 
PDH has the potential to achieve higher SNR and faster readout than conventional heterodyne or homodyne measurement.
However, a strong sideband tone could produce unwanted measurement-induced state transitions (MIST) during readout ~\cite{Blais:ionization, Blais:ionizationdynamics}. 
To quantitatively benchmark how much sideband amplitude the qubit can tolerate, we evaluate the state-transitions 
versus the frequency and power of a probe tone, using in a two-measurement scheme modified from Ref.~\cite{AngelaKouMIST}. 
Single-shot heterodyne measurements of the qubit state before and after application of the probe directly detect the qubit-state transitions induced.

The MIST measurement is configured in four steps. (i) A $\pi/2$ pulse is applied to the qubit to prepare an equal superposition of the ground and excited states. 
(ii) The first heterodyne measurement is performed, using the carrier frequency and $2~\mu$s duration chosen in Appendix~\ref{sec:stateassignment}. This measurement projects the qubit into one of the two computational basis states. 
(iii) After a wait time of $4~\mu$s (approximately $3\times$ the resonator lifetime), a probe signal with variable frequency and power is applied to the qubit for $2~\mu$s. 
(iv) After a second wait time of $4~\mu$s, another heterodyne measurement is performed (using the same configuration as the initial measurement).
The probe detuning $\Delta_p$ is varied from $2\pi \times 0$~MHz (the same as the heterodyne measurement tone) to $2\pi \times 40$~MHz below the carrier tone, spanning a range of approximately $66 \, \kappa$.
$5,000$ single-shot measurements are acquired for each combination of probe detuning and power. Reference measurements, as described in Appendix~\ref{sec:stateassignment}, are applied prior to measuring at new probe frequencies in order to compensate for long-time phase drift in the heterodyne readout.

Repeated application of an ideal QND measurement will always yield the same outcome, i.e. the conditional probability $p(j|j) = 1$, for any qubit state $\ket{j}$. However, the heterodyne measurements used here are not ideal and the qubit can experience decay during the wait time between the two measurements. Thus, non-zero transition probabilities $p(i|j)$ to find the qubit in state $\ket{i}$ in the second measurement and state $\ket{j}$ in the first will always be present. MIST due to the applied probe tone will manifest as increased transition probabilities above this baseline. In general for superconducting qubits, MIST takes the form of transitions to non-computational states \cite{google:MISTbeyondRWA, google:MISTwithinRWA}. Therefore, we quantitatively characterize MIST using both the transition probabilities between the computational states, $p(g|e)$ and $p(e|g)$, as well as the the transition probabilities to all other non-computational states, which we denote as $p(o|g)$ and $p(o|e)$.

When the probe detuning is zero, the possible range of applied powers is severely limited by the onset of added MIST, and significant increases in $p(o|g)$ and $p(o|e)$ are observed at approximately $-38$~dBm, indicating that heterodyne readout can only be carried out at carrier powers well below this level. However, as the probe detuning increases, the amount of MIST observed at fixed probe power decreases rapidly. Fig.~\ref{fig:detuningsweep_full} show sample first and second measurement IQ histograms for $-30$~dBm of probe power and probe detunings of $\Delta_p/2 \pi = 0,1,2,20$~MHz. At this power, a resonant probe ionizes the transmon almost completely, and the second measurement consists almost entirely of $\ket{o}$ events, but for larger detunings, no significant $\ket{o}$ population is observed.

For fixed probe detuning, there exists a threshold power above which probe-induced MIST dominates over all other imperfections in the system and the readout process breaks down. This threshold is lowest on resonance (approximately $-38$~dBm of applied power at room temperature) and increases with probe detuning. 
Figure \figqnd~in the main text shows conditional readout probabilities $p(i|j)$ from a representative subset of detunings, and the full dependence on pump power and detuning is shown in Fig.~\ref{fig:transitionprob_full}. 
For $\Delta_p \leq 2 \pi \times 5 \ \rm{MHz}\sim 8 \kappa$, a breakdown transition is clearly visible (Fig.~\ref{fig:transitionprob_full}(a-f)), and preliminary signs of breakdown are just visible at the highest power of $-10$~dBm when $\Delta_p = 2 \pi \times 10 \ \rm{MHz}\sim 16 \kappa$ (Fig.~\ref{fig:transitionprob_full}(g)). 
In contrast, for larger detunings no signitures of added probe-induced MIST are visible within the available dynamic range of probe powers (Fig.~\ref{fig:transitionprob_full}(h,i)). At these detunings, the qubit can tolerate at least $+28$~dBc of sideband power without adverse effects, corresponding to more than $14$~dB of heterodyne gain in a PDH signal.

\section{Practical Measurement Considerations for PDH Qubit Readout} \label{app:practical_consideration}

\subsection{Considerations for PDH measurement pulses}
One potential concern when adapting the PDH technique from optics to microwave qubit readout is the required quality of the phase modulation and the spectral purity of the generated sidebands. In optical PDH implementations, careful control of the modulation spectrum is essential because imperfections such as higher order sidebands lead to residual amplitude modulation which can degrade frequency discrimination \cite{JunYePDH2024}. In contrast, the requirements for qubit readout are not as strict.  In superconducting-qubit readout, the goal is simply to distinguish between different resonator responses corresponding to the two qubit states. Imperfections in the modulation spectrum may lead to an overall offset in the PDH signal but they do not prevent reliable quantum state discrimination.

Another consideration is that the PDH interrogation used here is pulsed rather than continuous-wave. The finite measurement window introduces potentially-deleterious spectral broadening of the interrogation tones.  However, this effect is also present in conventional pulsed heterodyne readout.  The nominal requirement is that the modulation frequency be sufficiently large compared to the inverse pulse duration so that the carrier and sidebands remain spectrally separated during the measurement window.  Superconducting-qubit experiments have already employed readout pulses with nontrivial structure, including shaped envelopes \cite{fastReadoutMartinis, mcclure2016rapid, readout2017Wallraff} in order to optimize signal acquisition and minimize measurement time. Thus, phase-modulated readout pulses do not constitute a significant additional complexity.

Lastly, as discussed in section~\ref{sec:MIST}, the PDH-specific requirement, is that the sidebands are sufficiently detuned from the carrier to minimize MIST.  

\subsection{Philosophy of measurement schemes and signal enhancement}

To clarify the role of PDH in signal enhancement (SNR, readout fidelities), it is useful to clearly distinguish three closely related measurement architectures. 

\subsubsection{Conventional heterodyne readout}

In conventional dispersive readout, a single microwave tone probes the resonator and the resulting signal is amplified and mixed with a local oscillator to extract the IQ components of the field. The measured signal depends directly on the absolute phase of the readout tone relative to the local oscillator, and therefore inherits the sensitivity to phase drift arising from generator instability and path-length fluctuations (see section~\ref{app:heterodyne_readout}). High-fidelity implementations of this approach almost universally employ parametric amplifiers (JPA~\cite{mutus2013JPA, yurke1989JPA, siddiqi2004JPA} or TWPA~\cite{ho2012TWPA}) to preserve the quantum-limited SNR during amplification.

\subsubsection{Synthetic PDH}

The measurements presented in this work use a synthetic PDH scheme in which the carrier and sidebands are simultaneously digitized using heterodyne detection, and the PDH signal is reconstructed digitally from the measured IQs (see section~\ref{sec:TripleDDC}.) In this configuration, the primary advantage is the improved phase stability from the intrinsic self-phase-referencing properties of PDH (section~\ref{app:insensitivity_of_pdh}). As we have shown, this leads to long-term phase stability at the sub-degree level even in the presence of hundred of degrees of absolute carrier phase drift. This method also has the advantage of having almost everything in common with conventional heterodyne readout, except the need to generate three input tones and carry out three digital downconversions, making implementation as straightforward as possible since it does not require custom detectors or output amplifier chains in the dilution refrigerator. However, because amplification occurs prior to the reconstruction of the PDH signal, in which the carrier and sidebands are multiplied together, this synthetic implementation does not directly benefit from the intrinsic heterodyne gain associated with full PDH detection using a square-law detector (see section \ref{sec:full_pdh}). 
In practice, increasing the sideband amplitudes reduces the noise that they contribute to the reconstructed PDH signal, but using very large sidebands is unlikely to be beneficial due to gain compression in the microwave parametric amplifier.

\subsubsection{Full PDH with square-law detection} \label{sec:full_pdh}

A third architecture, not implemented in the present experiment, is to detect the PDH beatnote directly with a cryogenic square-law detector. The PDH signal arises from the beatnote of the carrier and sidebands, and its amplitude scales as $E_{0}E_{\pm}$. As a result, strong sidebands sufficiently detuned from the carrier can potentially produce a heterodyne gain by generating a signal at the output of the cavity that is at least $14$~dB larger before entering the amplification chain. This increased signal amplitude can help preserve the quantum-limited SNR by reducing the impact of added noise in subsequent amplifiers and room-temperature electronics. We emphasize that the gain discussed in the main text does not corresponded to a $14$~dB improvement in the fundamental quantum-limited SNR. 

Achieving full PDH of this kind is of significant technical interest due to the potential to harness intrinsic heterodyne gain; however it requires development of a purpose-built Josephson-junction-based square-law detector.

\begin{figure}
    \centering
    \includegraphics[width=0.9\textwidth]{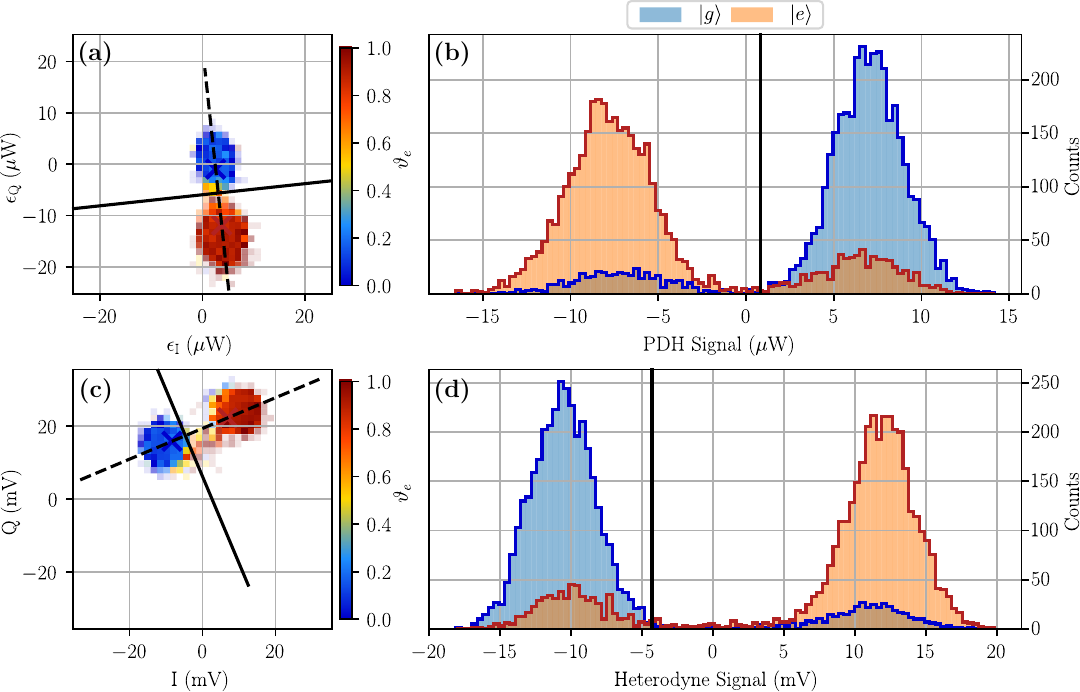}
    \vspace{-0.2cm}
    \caption{\textit{Heterodyne and reconstructed PDH single-shot readout fidelity.}
    (a) Single-shot IQ distribution for reconstructed PDH readout with the qubit prepared in $\ket{g}$ or $\ket{e}$. The dashed black line indicates the projection axis used to convert the two-dimensional distribution into the histogram in (b). Just as in \fref{fig:supple_carrierUnlocked_singleShot}, the color map indicates $\vartheta_{e}$, the fraction of shots corresponding to preparation in $\ket{e}$.  
    (b) Histogram of the projected PDH signal for the two states. The solid black line indicates the decision threshold chosen to maximize the readout fidelity, defined as $F = \frac{1}{2}\left[ p(g|g) + p(e|e) \right]$. The resulting single-shot fidelity, including state preparation errors, for the reconstructed PDH readout is $F = 86.44\%$.
    (c) Single-shot IQ distribution for conventional heterodyne readout using only the carrier signal from (a) and (b).
    (d) Histograms of the projected heterodyne signal for the two preparation states. The heterodyne readout fidelity is $F = 86.84\%$.
    Measurements were performed using a JPA with a carrier power of $-42$~dBm and the sidebands at $-45$~dBm with $30$~MHz detuning. The similar fidelity of the projected distributions illustrates that reconstructed PDH and conventional heterodyne readout achieve comparable single-shot discrimination under identical measurement conditions. 
    }
    \label{fig:HeterodyneVsPDH_Fidelity}
\end{figure}

\subsection{Measurement fidelity: heterodyne vs reconstructed PDH}
To illustrate the practical performance of the synthetic PDH and heterodyne readout, we use both methods to analyze the same single-shot dataset. The dataset consists of 5000 runs with the qubit prepared in $\ket{g}$ followed by a single measurement and 5000 runs with the qubit prepared in $\ket{e}$ prior to measurement. Since we are using the same dataset, both approaches employ the same linear amplification chain, JPA configuration, and measurement window, allowing a direct comparison of their readout fidelity. The resulting IQ distributions for heterodyne readout and the reconstructed PDH are shown in \fref{fig:HeterodyneVsPDH_Fidelity}. In this dataset the carrier power is $-42$~dBm, with both sidebands applied at $-45$~dBm and detuned by $30$~MHz from the carrier. Both techniques produce well-separated clusters corresponding to the ground and excited states, yielding comparable discrimination fidelity ($86.44\%$ for reconstructed PDH in \fref{fig:HeterodyneVsPDH_Fidelity}a-b, and $86.84\%$ for heterodyne readout in \fref{fig:HeterodyneVsPDH_Fidelity}c-d), under the present conditions. To quantify the state separation, we project the two-dimensional IQ distributions of the two states onto the axis of maximum separation between the two clusters, as indicated by the dashed black line in the IQ plots, producing the one-dimensional histograms shown in the adjacent panels. The reported fidelity values are calculated by applying a decision threshold that maximizes the readout fidelity $F = \frac{1}{2}\left[ p(g|g) + p(e|e) \right]$, which includes state preparation infidelity, and the chosen threshold is indicated by the solid black line in both the IQ plots and the histograms.

The fidelity values reported here should not be interpreted as a fundamental performance limit of either method. In particular, the device design and readout parameters used here were not optimized for maximum fidelity. Notably, the device does not incorporate Purcell filtering \cite{reedPurcellFilter}, the qubit coherence time is relatively short (see Table~\ref{tab:DeviceParameters}), and the qubit relaxation during the measurement window of $2\mu s$. Furthermore, the measured error rates in this section include contributions from state-preparation errors. Unlike the MIST measurement described in detail in section \ref{sec:MIST} , these data do not feature an initial measurement for projective state preparation. Importantly, the same readout parameters were used for both measurement methods and the comparison is intended to demonstrate that the reconstructed PDH readout achieves performance comparable to standard heterodyne detection in the same experimental configuration while providing substantially improved phase stability.

Finally, these measurements highlight a potential advantage of full PDH implementation using cryogenic square-law detection. In the present synthetic PDH approach, the signal is reconstructed from the three heterodyne tones and therefore inherits the technical noise of the heterodyne detection chain. In contrast, a full direct PDH detection scheme would generate the beatnote signal prior to amplification, allowing the intrinsic heterodyne gain to produce a larger signal at the input of the amplification chain, and reduce the impact of downstream noise from classical amplifiers. 

\section{Device Description and Wiring Diagram}\label{app:deviceandwiring}
\begin{figure}[t!]
    \centering
    \includegraphics[width=0.48\textwidth]{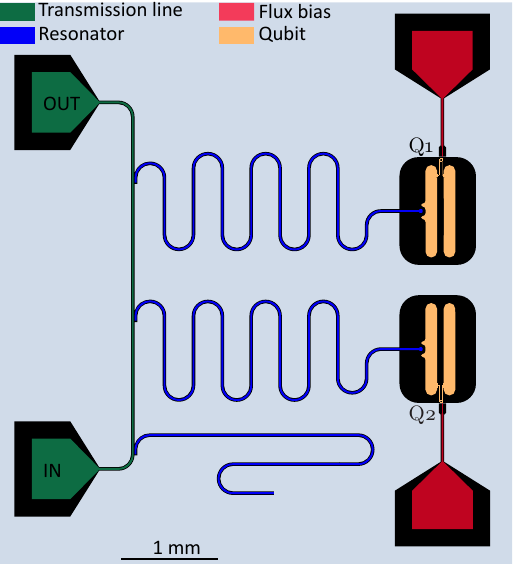}
    \vspace{-0.2cm}
    \caption{\textit{Device Design.}
    The device consists of three hanger-coupled resonators and two flux-tunable transmon qubits. Each qubit is flux-biased with on-chip flux-bias line and labeled with its name. The resonators are coupled to both the transmission line and the qubits for drive and readout purposes. The third bare resonator allows measurement of the PDH response independent of MIST in a qubit. All measurements reported in this letter are from $Q1$. The corresponding parameters are listed in the Table~\ref{tab:DeviceParameters}.
    }
    \label{fig:Devicedesign}
\end{figure}

\begin{figure*}[t]
    \centering
    \includegraphics[width=\textwidth]{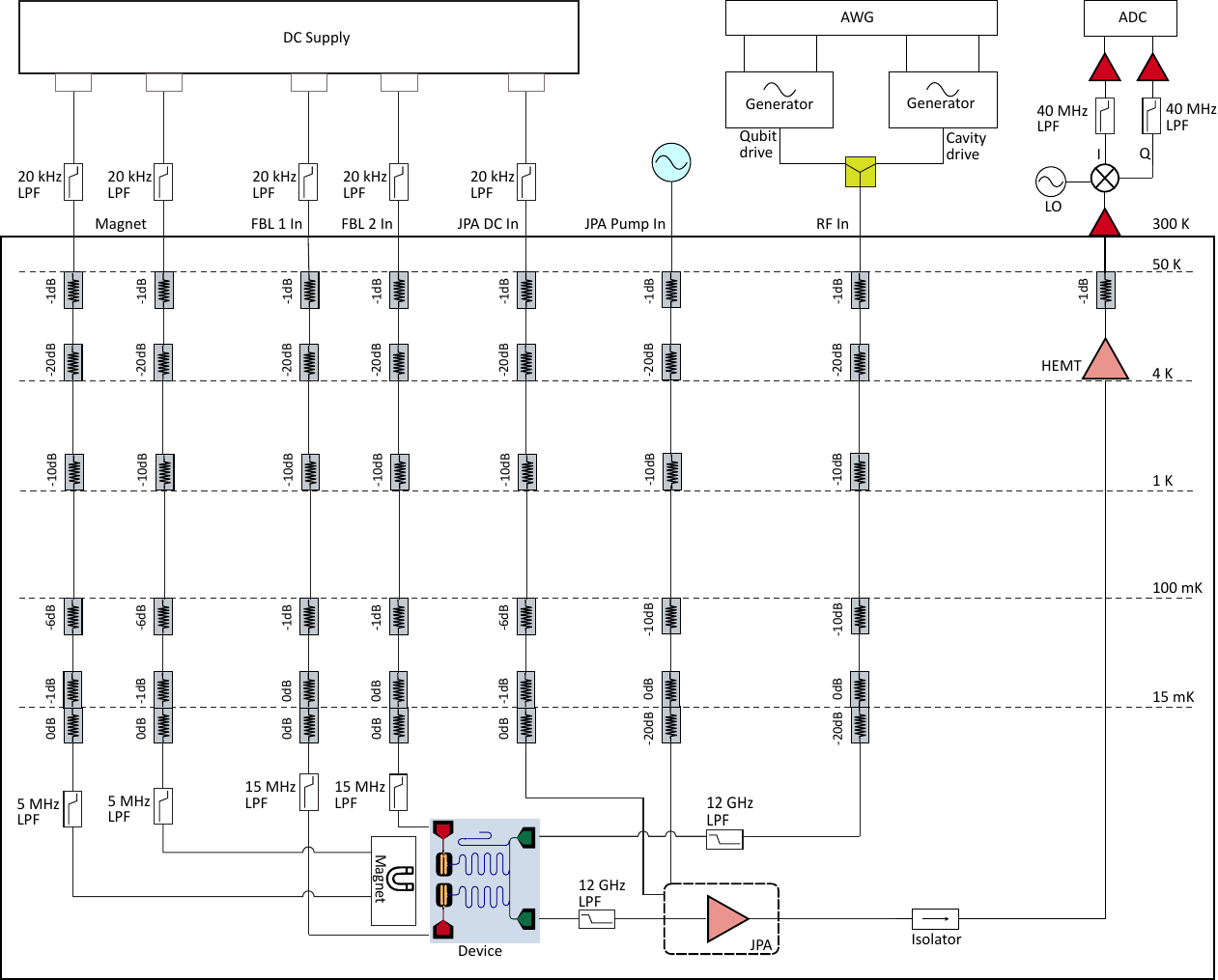}
    \caption{\textit{Schematic of the measurement setup for the PDH readout scheme.}
    A four-channel arbitrary waveform generators (Zurich Instruments HDAWG) and two vector signal generators (Rohde-Schwarz SGS100A) produce the PDH carrier/sidebands and qubit-drive tones. The combined signals are attenuated before reaching the device. The input and output lines are filtered and isolated to prevent noise and back-action. The outgoing signal is amplified by a JPA (RTX BBN WB-JPA) and a HEMT amplifier (LNF-LNC4\_8G), then mixed with a local oscillator for heterodyne detection. A digitizer (ADC) records the heterodyne signals from all three tones simultaneously. Flux biasing is provided by two on-chip lines and an external solenoid magnet (symbol: U magnet), each filtered to suppress unwanted AC components.
    }
    \label{fig:wiringDiagram}
\end{figure*}

The device was fabricated on a $7\times7$~mm$^{2}$ sapphire wafer coated with a 200-nm-thick tantalum film deposited by StarCryo. The CAD layout of the device is shown in Fig.~\ref{fig:Devicedesign}. The device consists of three coplanar waveguide (CPW) resonators (blue), hanger-coupled to a feed line (green). The S$_{21}$ transmission measurements of the hanger geometry are effectively equivalent to measuring reflection in an optical PDH scheme. As a result, a large sideband detuning relative to the resonator linewidth can be used, and no circulator is required in this measurement configuration. The Josephson junctions in the two flux-tunable transmon qubits (orange) are Al–AlO$_\mathrm{x}$–Al junctions patterned using electron-beam lithography (Elionix-G100) and deposited using double-angle evaporation (Plassys MEB 550S). An external solenoid magnet, shown in the wiring diagram in Fig.~\ref{fig:wiringDiagram}, is used to tune the qubit operating frequencies via flux bias. In addition, two on-chip flux-bias lines (red)—one for each qubit—provide independent and precise frequency control. Although the device contains two qubits, all measurements reported here were performed using only $Q1$, the upper qubit. The relevant parameters of this qubit are summarized in Table~\ref{tab:DeviceParameters}.

\begin{table*}[t]
    \caption{\textit{Device parameters}}
    \label{tab:DeviceParameters}
    \begin{tabular}{||c|c|c||}
        \hline
        Parameter  & Symbol & Value \\
        \hline \hline
        Qubit g-e frequency  & $\omega_{ge}/2\pi$ (GHz) & 5.968 \\
        Qubit anharmonicity & $\alpha/2\pi$ (MHz) & -156.5 \\
        Qubit relaxation time & $T_{1}$ $(\mu s)$ & 21\\
        Qubit Ramsey time & $T_{2}^{\text{R}}$ $(\mu s)$& 3.876 \\
        \hline
        Resonator frequency when qubit is at $\ket{g}$ & $\omega_{r,\ket{g}}/2\pi$  (GHz) & 6.773 \\
        Dispersive shift & $2\chi/2\pi = f_{r, \ket{e}} - f_{r, \ket{g}}$ (MHz) & -1.189\\
        Resonator linewidth & $\kappa/2\pi$ (kHz) & 594\\
        \hline
    \end{tabular}   
\end{table*}

The PDH readout scheme is measured utilizing an arbitrary waveform generator (AWG, Zurich Instruments HDAWG) to generate the desired sidebands simultaneously with the carrier tone via IQ modulation of a vector signal generator (Rohde-Schwarz SGS100A). A second IQ modulated vector signal generator is used to generate the qubit drive tone. The two signals are combined using a power splitter (ZFRSC-183-S+) and undergo $-60$~dB attenuation in the dilution refrigerator. The input and output of the device
are protected by two identical low pass filters (LPF, K\&L 6L250-00089) with a cut-off frequency of 12 GHz. Additionally, the device is protected by an isolator (QCI-G0401202AS), preventing thermal radiation in the output lines from reaching the device. At room temperature, the output signal is mixed with the local oscillator and downconverted prior to digitization. The digitizer card, or the analog-to-digital convertor, (ADC, Acqiris SA220P-1013), records the heterodyne signal for each of the three PDH tones, and their IQ components are extracted by digital downconversion. 

In the absence of an optimized cryogenic $|E^2|$ detector, we use a conventional microwave-band linear amplification chain to extract readout signals from the device, consisting of a Josephson parametric amplifier (JPA, RTX BBN WB-JPA) and a high electron mobility transistor amplifier (HEMT, LNF-LNC4\_8G). Since each of the three PDH tones is extracted and detected separately, rather than directly beating the carrier and sidebands together and accessing the intrinsic heterodyne gain of conventional PDH, the low-noise amplification of the JPA is critical for achieving single-shot readout. However, the limited bandwidth and asymmetry of the gain profile of the JPA introduce an imbalance between the sidebands, which can be seen clearly in Fig.~\ref{fig:supple_carrierUnlocked_singleShot}. This asymmetry can be removed via precompensation, as mentioned in Appendix~\ref{sec:phasestability}. However, it does not prevent single-shot qubit readout and is not intrinsic to PDH readout of superconducting qubits. Future implementation of direct PDH with a cryogenic $|E^{2}|$ detector would not involve a microwave-band parametric amplifier, and large scale implementations of synthetic PDH on multiple qubits would use a wideband traveling-wave parametric amplifier (TWPA) \cite{quantumengineersguide}, rather than a narrowband JPA like the one used here. Non single-shot measurements are taken with the JPA deactivated, in Fig.~\ref{fig:supple_readoutSchematics_ModFreqSweep}, to show the fundamental response of the readout resonator and input/output lines.

\bibliographystyle{apsrev4-2}
\bibliography{refs.bib}